\documentclass[a4paper,fleqn]{cas-dc}

\usepackage[numbers]{natbib}

\usepackage{amssymb}
\usepackage{graphicx}
\usepackage{url}
\graphicspath{ {./images/} }
\usepackage{soul}
\usepackage{changepage}
\usepackage{hyperref}
\usepackage{graphicx}
\usepackage{subcaption}
\usepackage{balance}
\usepackage{colortbl}
\usepackage{tabularx}
\usepackage{tabu}
\usepackage{booktabs}
\usepackage{xparse}
\usepackage{xspace}
\usepackage{multirow}
\usepackage{enumitem}
\usepackage{adjustbox}
\usepackage{verbatim}
\usepackage{makecell, cellspace, caption}
\usepackage{tikz}
\usepackage{xcolor}

\newlength\MAX  

\newcommand*\ChartPWDi[1]{#1~\rlap{\textcolor{blue!20}{\rule{\MAX}{2ex}}}\textcolor{blue}{\rule{#1\MAX}{2ex}}}
\newcommand*\ChartPW[1]{#1~\rlap{\textcolor{RedOrange!20}{\rule{\MAX}{2ex}}}\textcolor{RedOrange}{\rule{#1\MAX}{2ex}}}
\newcommand*\ChartDi[1]{#1~\rlap{\textcolor{OliveGreen!20}{\rule{\MAX}{2ex}}}\textcolor{OliveGreen}{\rule{#1\MAX}{2ex}}}
\newcommand*\ChartNone[1]{#1~\rlap{\textcolor{red!20}{\rule{\MAX}{2ex}}}\textcolor{red}{\rule{#1\MAX}{2ex}}}
\def\tsc#1{\csdef{#1}{\textsc{\lowercase{#1}}\xspace}}
\tsc{WGM}
\tsc{QE}
\tsc{EP}
\tsc{PMS}
\tsc{BEC}
\tsc{DE}

\definecolor{pblue}{rgb}{0.13,0.13,1}
\definecolor{pgreen}{rgb}{0,0.5,0}
\definecolor{pred}{rgb}{0.9,0,0}
\definecolor{pgrey}{rgb}{0.46,0.45,0.48}
\definecolor{codebackground}{rgb}{0.95, 0.95, 0.92}
\definecolor{gray50}{gray}{.5}
\definecolor{gray40}{gray}{.6}
\definecolor{gray30}{gray}{.7}
\definecolor{gray20}{gray}{.8}
\definecolor{gray10}{gray}{.9}
\definecolor{gray05}{gray}{.95}
\definecolor{arsenic}{rgb}{0.23, 0.27, 0.29}
\usepackage{tikz}

\renewcommand\toprule{\specialrule{1pt}{1pt}{0pt}\rowcolor{gray!20}}
\renewcommand\midrule{\specialrule{0.4pt}{0pt}{0pt}}

\RequirePackage{fontawesome}
\newcommand{\keyfindingsone}[1]{ %
    \vspace{5pt} %
    \noindent\fcolorbox{arsenic}{gray10}{%
        \parbox{1\linewidth}{%
          \textbf{} #1 %
        }%
    }%
    \vspace{5pt} %
}%

\begin{document}
\let\WriteBookmarks\relax
\def\floatpagepagefraction{1}
\def\textpagefraction{.001}

\shorttitle{Detecting Privacy Requirements from User Stories}

\shortauthors{Francesco Casillo et~al.}

\title [mode = title]{Detecting Privacy Requirements from User Stories with NLP Transfer Learning Models}                      
\tnotemark[1]

\tnotetext[1]{This work has been partially supported by the Italian Ministry of Education, University and Research (MIUR) under grant PRIN 2017 ``EMPATHY: Empowering People in deAling with internet of THings ecosYstems'' (Progetti di Rilevante Interesse Nazionale - Bando 2017, Grant 2017MX9T7H).}

%
\author[1]{Francesco Casillo}[orcid=0000-0003-4869-8068]
\ead{fcasillo@unisa.it}
\credit{Methodology, Software, Validation, Writing - Original Draft}
\address[1]{Department of Computer Science, University of Salerno, Via Giovanni Paolo II, 132, Fisciano(SA), 84084, Italy}

\author[1]{Vincenzo Deufemia}[orcid=0000-0002-6711-3590]
\ead{deufemia@unisa.it}
\credit{Conceptualization, Formal analysis, Writing - Review \& Editing, Supervision}

\author[1]{Carmine Gravino}[orcid=0000-0002-4394-9035]
\ead{gravino@unisa.it}

\credit{Conceptualization, Formal analysis, Writing - Review \& Editing, Supervision}

\cormark[1]

\cortext[cor1]{Corresponding author}

\begin{abstract}
\  \textbf{Context:}
To provide privacy-aware software systems, it is crucial to consider privacy from the very beginning of the development.
However, developers do not have the expertise and the knowledge required to embed the legal and social requirements for data protection into software systems.\\
\textbf{Objective:}
We present an approach to decrease privacy risks during agile software development by automatically detecting privacy-related information in the context of user story requirements, a prominent notation in agile Requirement Engineering (RE).\\ 
\textbf{Methods:}
The proposed approach combines Natural Language Processing (NLP) and linguistic resources with deep learning algorithms to identify privacy aspects into User Stories. NLP technologies are used to extract information regarding the semantic and syntactic structure of the text. This information is then processed by a pre-trained convolutional neural network, which paved the way for the implementation of a Transfer Learning technique. We evaluate the proposed approach by performing an empirical study with a dataset of 1680 user stories. \\
\textbf{Results:}
The experimental results show that deep learning algorithms allow to obtain better predictions than those achieved with conventional (shallow) machine learning methods. Moreover, the application of Transfer Learning allows to considerably improve the accuracy of the predictions, ca. 10\%.\\
\textbf{Conclusions:} Our study contributes to encourage software engineering researchers in considering the opportunities to automate privacy detection in  the early phase of design, by also exploiting transfer learning models.
\end{abstract}




\begin{keywords}
User Stories \sep Natural Language Processing \sep Deep Learning \sep Transfer Learning
\end{keywords}

\maketitle

\section{Introduction}
\label{sec:sample1}

Requirements engineering (RE) is one of the most complex activity of software engineering. Misunderstandings and imperfections in the requirement documents can easily lead to design flaws and cause several problems \cite{Sommerville97RE,10.5555/1869735}. 
Agile RE is based on face-to-face collaboration between customers and developers which helps to address several RE problems, but this does not exclude the presence of others. 
Among them, the detection of non-functional requirements (NFRs) by stakeholders is often a difficult activity due to several reasons \cite{twenty}. 
To alleviate this problem, several solutions for the automatic detection of NFRs from text documents have been proposed  \cite{DBLP:conf/wetice/PaetschEM03, RE2017, DBLP:conf/icse/Nguyen09,6611715}. For instance, Slankas \textit{et al.} have proposed a tool-based approach, named \textit{NFR Locator}, to extract sentences in unconstrained natural language documents, which are classified into one of the 14 defined NFR categories \cite{6611715}. 
In general, these NFR detection tools provide only an overview of the identified NFRs. However, since stakeholders usually have expertise in few specific areas, they might have difficulties in defining all the features of a software application, increasing the risk of neglecting some of them \cite{twenty}.

Privacy is an essential NFR that needs special attention as business needs require data protection and safeguarding \cite{7961663}. Even if privacy requirements frequently appear in software documentations, most of the time stakeholders ignore them.
The difficulty of privacy requirement identification mainly depends from the quality of requirement specifications as shown in several studies (e.g., \cite{nine, thirtyfour}). 

In this paper we propose a deep learning approach to identify possible privacy requirements within User Stories (USs). 
The proposed solution aims to support  practitioners, with poor privacy expertise, in the identification of NFRs related to privacy.
Although a lot has been done in the field of privacy detection, to the best of our knowledge no study deals with the analysis of USs. Thus, we verify whether it is  possible to exploit knowledge and tools proposed to address similar problems. With respect to conventional machine learning methods, the deep ones have unique advantages in feature extraction and semantic mining \cite{lecun2015deeplearning}, and have achieved excellent results in text classification tasks \cite{10.1007/978-3-030-21373-2_14, DBLP:journals/corr/abs-2008-00364, HANECZOK2020102371, 10.1093/jamia/ocz149, 8945891, 7930091}. 
Thus, from the analysis of user stories the deep learning models can infer individual privacy information and privacy rules, which  can be used to recognize privacy-related entities for individual user stories. Then, the users can be reminded of the possibility of privacy leakage, based on the defined privacy rules.

The proposed approach combines the use of linguistic resources and Natural Language Processing (NLP) techniques to extract features useful not only to capture the semantic meaning and the syntactic structure of the text, but also to determine the presence or absence of privacy-related words. 
A further peculiarity of our approach is the use of Transfer Learning (TL), an emergent strategy where a system developed for a task is reused for a model on a different but related task \cite{torrey2010transfer,DBLP:journals/ese/KocaguneliMM15,DBLP:journals/tse/KrishnaM19}. 
Specifically, we use a pre-trained convolutional neural network (CNN) designed to identify  personal, private disclosures from short texts \cite{10.1007/978-3-030-21373-2_14} to extract features from user stories, which are combined with features obtained from a privacy dictionary to construct a US-privacy classifier.

To show the effectiveness of our approach, we present the results of an empirical study carried out by exploiting  a dataset of 1680 user stories taken from \cite{Dalpiaz2018RequirementsDS}. %
In particular, we  present  a  type  of  sanity  check by  formulating two research questions with the aim of verifying if a deep learning method (CNN) performs at least as conventional (shallow) machine learning  methods, 
when exploiting NLP-based features (\textbf{RQ1}) or privacy word features (\textbf{RQ2}). 
The sanity check allows us to verify whether the further effort needed to apply CNN is payed back by an improvement in the prediction accuracy, and the possible contribution of PW features when applying  shallow and deep learning  methods.

The comparison between shallow and deep learning methods is often performed when evaluating  text classification tools (e.g., \cite{DBLP:journals/corr/abs-2008-00364, HANECZOK2020102371, 10.1093/jamia/ocz149}), mainly due to the possible noise in the data that can lead to substantial changes in the accuracy of decisions \cite{DBLP:journals/corr/abs-2008-00364}. 
In particular, in some studies, shallow learning methods outperformed the deep ones in text classification tasks \cite{10.1093/jamia/ocz149}.
In our study, no clear result is obtained in the comparison when exploiting PW features (RQ2), thus confirming the importance of performing this kind of check. Differently, 
the results about RQ1 show that the deep learning method performs significantly better than the conventional machine learning  methods, when exploiting NLP-based features.

After performing the two sanity checks, we investigate the proposed NLP-based Transfer  Learning  method by formulating a third research question (\textbf{RQ3}) aiming to compare its performances with those achieved using deep learning methods based on NLP-based features or privacy word (PW) features. 
The experimental results for RQ3 reveal an improvement of more than 10\% (in terms of both Accuracy and F1-score \cite{Baeza-Yates:1999}) compared to the individual CNNs.

\textit{Organization of the paper.}
Section \ref{sec:related} presents the research background on agile requirement engineering and how privacy is typically analyzed in this context. Section \ref{sec:method} describes the approach designed to identify privacy aspects in agile requirement specifications. Section \ref{sec:studydesign} describes the design of the empirical study carried out to evaluate the approach. Section \ref{sec:results} reports on quantitative results and discusses the main findings. 
Section \ref{sec:conclusions} concludes the paper and presents future research directions.

\section{Related work}
\label{sec:related}

This section analyzes the different NLP techniques proposed in literature for US analysis. USs typically follow a structured format characterized by the \textit{who}, the \textit{what}, and the \textit{why} of a requirement, becoming a standard de facto  \cite{formatus}. 
An example of user story defined by using the Cohn's model \cite{cohnmodel} is:

\keyfindingsone{\texttt{As a site member, I want to access to the Facebook profiles of other members so that I can share my experiences with them}}

Several frameworks and methodologies have been proposed for analyzing the quality of USs through their syntactic analysis, with the aim of making them more accurate and clear for customer's requirement definition \cite{9105527,7320415,heck2014quality}.
Other relevant investigations concern with the transformation of USs into models and components useful for the next stages of the software development processes. In particular, software diagrams can be automatically generated from USs in order to provide a visual representation for project stakeholders, to identify conceptual entities, or to highlight potential problems in US definition \cite{DBLP:journals/spe/KaraaASDAG16,Procedia,Procedia2,DBLP:journals/re/LucassenRDWB17}. These activities open the door to further automated analysis able to generate conceptual models \cite{7765525}, Use Case scenarios \cite{8587287}, and even Backlog Items \cite{Mter2019RefinementOU}.
USs can also be analyzed to automatically generate test cases \cite{Rane2017AutomaticGO} and create behavioral models that help testers who might be non-expert.

Other studies on USs focus on extracting attributes that can guide architecture design without relying on systematically and formally defined knowledge. For example, Gilson \textit{et al.} show that USs  might have a great impact on early stage decisions (because they might implicitly refer to quality attributes) allowing software architects to have an idea of the consequences of the possible design decisions \cite{8712367}. In particular, they use machine learning (ML) techniques to classify if the USs refer to quality attributes and, if so, which ones they refer to. 

Approaches dealing with other types of attributes are able to provide more detailed information, which improves the definition of customer requirements and facilitates decisions made during the software development process \cite{twenty}. For example, Villamizar \textit{et al.} define an approach for reviewing security-related aspects in agile requirement specifications with a focus on web applications \cite{8920644}. Their results indicate significant differences when comparing the performance achieved by experts using their approach against other defect-based techniques. Similarly, Riaz \textit{et al.} propose a ML-based tool that takes in input a set of natural language artifacts and automatically identifies (and classifies) security-relevant phrases according to predefined security objectives \cite{6912260}. However, stakeholders may not be able to assess and define all aspects of a software application together with customers, increasing the risk of leaving out even high priority ones \cite{twenty}, as in the case of data privacy. 

Many efforts have been devoted to privacy disclosure in the recent years, both to facilitate the work of analysts and developers \cite{10.1007/978-3-642-02843-4_7,Vimercati12dataprivacy} and to define a linguistic taxonomy of privacy for content analysis \cite{10.1145/1978942.1979421,PrivacyDictionary}. Many of the privacy detection approaches focus on the automatic recognition of sensitive personal information in unstructured text \cite{8945891,9162683,10.1007/978-3-030-21373-2_14}, which allows to develop several interesting tools, such as TABOO \cite{7930091} and PrivacyBot \cite{8931855}. On the other hand, many companies have particular needs with respect to personal data processing, and in the software design phase these needs may be set aside to make space for more functional requirements \cite{10.1145/2568225.2568244}. Therefore, the identification of privacy content can be considered crucial when building the architecture of a software system. However, to the best of our knowledge, nothing is proposed in the literature about the automatic identification of privacy content in the early stages of Agile software development. This work aims to fill this gap, by providing and evaluating an approach for detecting privacy information from USs.

\section{A Methodology for Privacy Disclosure Detection within User Stories}
\label{sec:method}
The proposed technique aims at identifying privacy-related threats in agile requirement specifications. The approach considers USs and linguistic resources as input, and exploits NLP techniques to determine the presence or absence of privacy-related words in the USs. The latter are structured in a sentence as follows \cite{8920644}:
\begin{center} 
\texttt{As a} [\textit{role}]\texttt{, I want to} [\textit{feature}]\texttt{, so that} [\textit{reason}]. 
\end{center}
\noindent
Although this structure simplifies US's comprehension, 
the detection of privacy disclosures may be ineffective due to a wide variety of possible terms in USs.
Therefore, more advanced approaches are needed to improve its effectiveness. 

The proposed method leverages convolutional deep neural networks to identify short texts of USs having private disclosures. 
In particular, we first adopt a lexicon-based approach to identify the words having entity-level privacy disclosures, by using the matches between USs and a privacy dictionary as machine learning features. This method can give high precision, but low recall since it relies only on the count of sensitive words in a document, without considering the context in which these words are used.
To improve recall, we also exploit NLP tools to derive linguistic features, such as syntactic dependencies and entity relations, which keep the sentence level context into consideration.

The proposed deep neural network model combines together multiple channels to perform the disclosure/non-disclosure classification task. Each channel refers to different representations of the same candidate user story. 

To deal with the paucity of curated data in the field, we propose the use of transfer learning, which allows to utilize knowledge acquired for one task to solve related ones. In particular, we exploit a pre-trained CNN that exploits NLP-based features to detect privacy disclosures in Reddit users’ posts and comments. This neural network is trained on 10K disclosure and non-disclosure sentences. 

In what follows we provide details of the features used for privacy disclosure detection (Sections \ref{3.1} and \ref{3.2}), and the architectures of the considered deep neural network models (Sections \ref{3.3} and \ref{3.4}).

\subsection{Lexicon-based privacy disclosure features}
\label{3.1}

These features are extracted from the text of the USs by using linguistic resources, i.e., dictionaries, containing individual words or phrases that are assigned to one or more linguistic categories. By using a privacy dictionary it is possible to count the occurrences of each dictionary word within a US text, incrementing the relevant categories to which the words belong \cite{10.1145/1978942.1979421,PrivacyDictionary}. The final result consists of values for each linguistic privacy category, represented as a percentage of the total words in the text.

We use the privacy dictionary proposed by Vasalou \textit{et al.} \cite{PrivacyDictionary}, which constructed and validated eight dictionary categories on empirical material from a wide range of privacy-sensitive contexts. Experimental results have shown that the identified categories allow to effectively detect privacy language patterns within a given text.
Figures \ref{fig:privawordscat_example1} highlights the two US words  contained in the privacy dictionary defined in \cite{10.1145/1978942.1979421,PrivacyDictionary}, whereas Table  \ref{fig:privawordscat_example2} reports the information on the privacy category \textit{OpenVisible} they belong to.

\begin{figure*}[width=\textwidth, ht]
\centering
\noindent\fcolorbox{arsenic}{white}{\textbf{\footnotesize{As a site member, I want to \colorbox{red!90}{access} to the Facebook profiles of other members so that I can \colorbox{red!90}{share} my experiences with them}}}
\caption{The US words highlighted in red are contained in the \textit{privacy dictionary} defined in \cite{PrivacyDictionary}.}
\label{fig:privawordscat_example1}
\end{figure*}

\begin{table*}[width=\textwidth, htb]
\centering
\begin{tabular*}{\tblwidth}
{@{} LLL@{} }
\toprule
\rowcolor{gray!0}
\textbf{Category name (number of words)} & \textbf{Description} & \textbf{Example dictionary words} \\
\midrule
\textit{OpenVisible ( 2 )} &  \textit{open and public access to people} & \textit{port, display, accessible} \\
\bottomrule
\end{tabular*}
\caption{The privacy category of the words `access' and `share' \cite{10.1145/1978942.1979421}.}
\label{fig:privawordscat_example2}
\end{table*}

\subsection{NLP-based features for privacy disclosure}
\label{3.2}

These linguistic features are obtained from the text of the USs by extrapolating entities, the parts of speech, and the dependencies between them, since the aim is to understand the text from its meaning and to capture those features that may affect the classification of  USs as related to privacy disclosure. 

In what follows, we describe how these NLP-based features have been obtained for the US shown in Figure \ref{fig:privawordscat_example1}
by using the NLP spaCy toolkit\footnote[1]{\url{https://spacy.io/}}.
First, the text is pre-processed by removing punctuation and insignificant words, leaving only lexical items (\textit{tokenization}). The result of this process for the considered US is:
[`As', `a', `site', `member', `I', `want', `to', `access', `to', `the', `Facebook', `profiles', `of', `other', `members', `so', `that', `I', `can', `share', `my', `experiences', `with', `them']

The Dependency Parser (DP) Toolkit\footnote[2]{\url{https://spacy.io/usage/linguistic-features\#dependency-parse}} from spaCy has been used to extract information on syntactic relations and parts of speech (POS), which enable the data to be enriched with such information on syntactic and semantic structure. Table \ref{table:table_pos_dep} reports the POSs and dependencies extracted from the user story of Figure \ref{fig:privawordscat_example1}. 
These features help the model to understand the common sequence of tokens and the occurrence of dependency tags \cite{evans-zhai-1996-noun}. 

\begin{table}[width=1\linewidth,cols=3,pos=h]
\caption{Parts-of-speech and dependencies extracted from a user story.}\label{table:table_pos_dep}
\begin{tabular*}{\tblwidth}{@{} CCC@{}}
\toprule
\rowcolor{gray!0}
\textbf{Text} & \textbf{Part of speech} & \textbf{Dependency}\\
\midrule
As & SCONJ&prep\\
a & DET&det\\
site & NOUN&compound\\
member & NOUN&pobj\\
I & PRON&nsubj\\
want & VERB&ROOT\\
to & PART&aux\\
access & VERB&xcomp\\
to & ADP&prep\\
the & DET&det\\
Facebook & PROPN&compound\\
profiles & NOUN&pobj\\
of & ADP&prep\\
other & ADJ&amod\\
members & NOUN&pobj\\
so & SCONJ&mark\\
that & SCONJ&mark\\
I & PRON&nsubj\\
can & VERB&aux\\
share & VERB&advcl\\
my & DET&poss\\
experiences & NOUN&dobj\\
with & ADP&prep\\
them & PRON&pobj\\
\bottomrule
\end{tabular*}
\end{table}

The Named Entity Recognizer (NER)\footnote[3]{\url{https://spacy.io/usage/linguistic-features\#named-entities}} of spaCy has been used to assign labels to contiguous tokens. The default model provided by the library identifies various entities, such as companies, locations, organizations, and products, and new entities can be added to the system by updating the model with new data.

\subsection{Deep Neural Network Models}
\label{3.3}

\begin{figure*}[width=\textwidth, htb]
\centering
\includegraphics[width=1\textwidth]{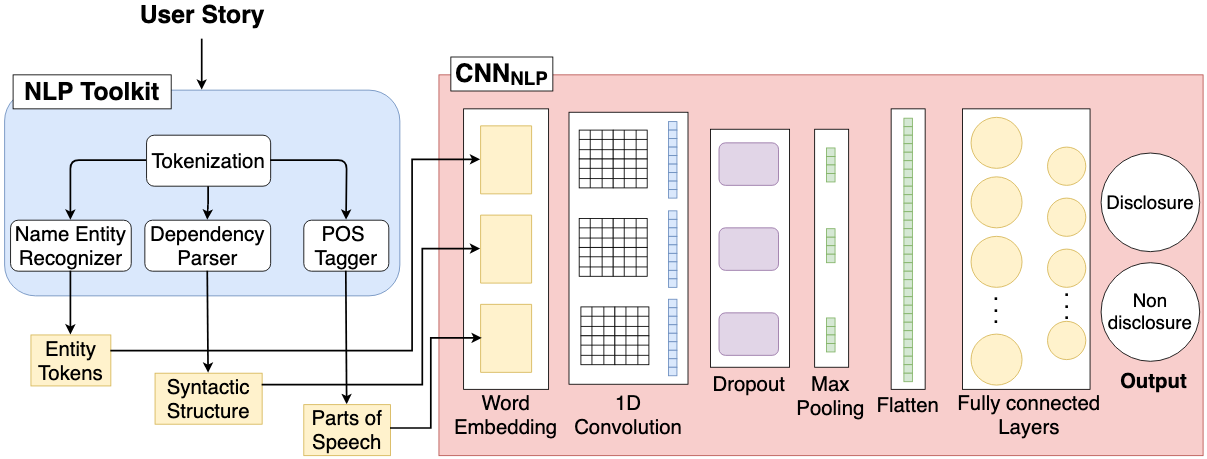}
\caption{The $CNN_{NLP}$ architecture.}
\label{fig:cnn_nlp}
\end{figure*}

\begin{figure*}[width=\textwidth, htb]
\centering
\includegraphics[width=1\textwidth]{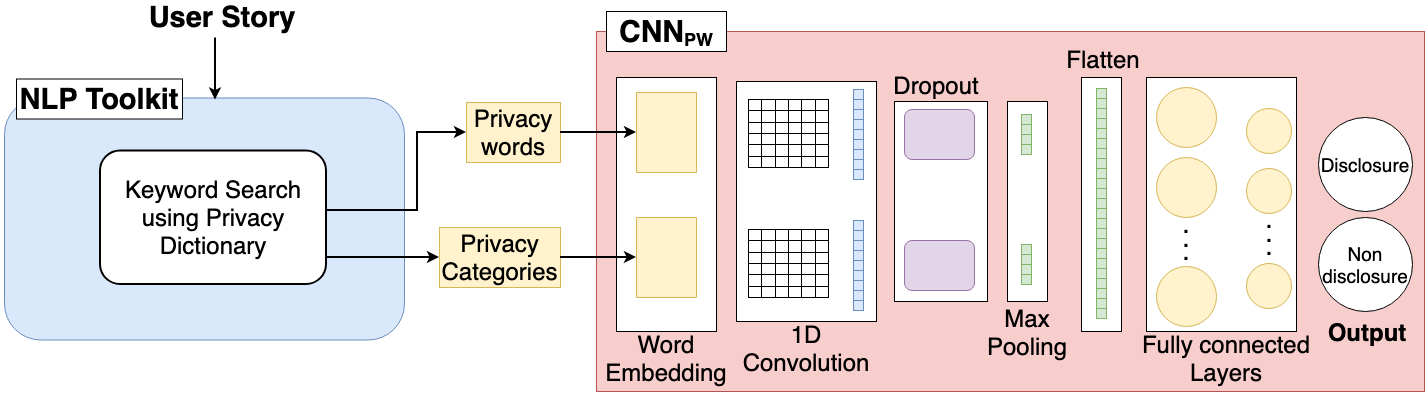}
\caption{The $CNN_{PW}$ architecture.}
\label{fig:cnn_pw}
\end{figure*}

After doing all the necessary pre-processing steps, the data is then fed into a multi-input deep neural network to learn the hidden patterns and features to distinguish between texts having disclosure and non-disclosure occurrences. In particular, we constructed two deep convolutional neural networks, one based on the NLP-based features introduced in Section \ref{3.2} (see Figure \ref{fig:cnn_nlp}), the other on the lexicon-based features of Section \ref{3.1} (see Figure \ref{fig:cnn_pw}). 
The first takes lexical (word tokens) features through one input, syntactical features (dependency parse tree information) through another input following a merging of those feature vectors. Later these vectors additionally get merged with supplemental (auxiliary) inputs before going through a further multi-layer perceptron stage. At the end of the deep neural network, a single neuron is used to provide the probability toward each of the above mentioned classes.
The latter performs similar operations fed the features obtained from the privacy dictionary.

\subsection{A Transfer Learning Methodology for Privacy Disclosure Detection}
\label{3.4} 
The previous deep neural networks require a specific dataset to train the model from scratch to the specific classification task. Unfortunately, the datasets of user stories available in literature contain few hundreds of examples. For this reason, it could be possible that the models are not able to adequately learn how to classify a US. To deal with this problem we introduce a neural network model that exploits transfer learning for the disclosure/non-disclosure classification task.

Transfer learning is an approach in which the knowledge learned from a large-scale dataset to solve a particular task is reused (transferred) and applied to solve a different but related task \cite{torrey2010transfer}. In particular, transfer learning allows to use pre-trained shallow/deep learning models by fine-tuning them on a relatively small labeled dataset from the downstream task.

In the proposed model, the NLP-based features described in Section \ref{3.2} are processed by a pre-trained convolutional neural network whose aim is to identify short texts that have personal, private disclosures \cite{10.1007/978-3-030-21373-2_14}.  
In particular, the neural network identifies whether the unstructured text given as input contains private disclosures by analyzing the semantic and syntactic structure of the text through the extraction of the characteristics described above, i.e., entities, dependencies, and parts of speech. 
This network has been trained for privacy disclosure classification on ten thousand Reddit  users’ posts and comments.

\begin{figure*}[width=\textwidth, htb]
\centering
\includegraphics[width=0.75\textwidth]{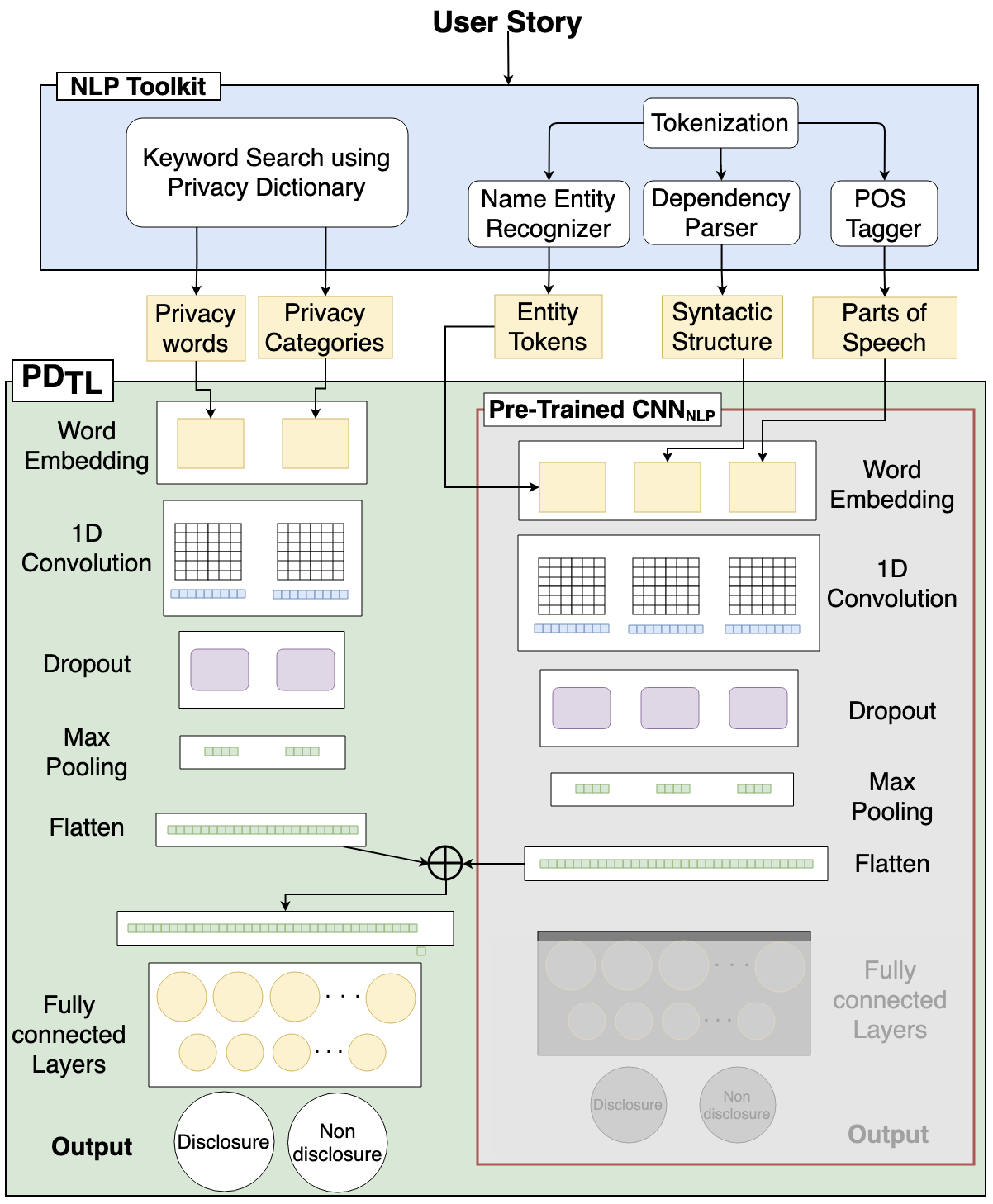}
\caption{The $PD_{TL}$ architecture.}
\label{fig:CNN_TD}
\end{figure*}

Figure \ref{fig:CNN_TD} shows the architecture of the deep neural network, named $PD_{TL}$, obtained by applying transfer learning. 
Taking advantage of the flexibility of the tools provided by Keras\footnote[1]{\url{https://keras.io/}}, the pre-trained neural network proposed in \cite{10.1007/978-3-030-21373-2_14} has been truncated after the Flatten layer. The latter is concatenated with the Flatten layer of another neural network that processes the lexicon-based privacy features. As a consequence, the resulting neural network processes the information concerning the semantic and syntactic structure, enriching this analysis with the information derived from the privacy dictionary.

\section{Empirical study design}
\label{sec:studydesign}
In this section we present the design of the empirical study we have performed. In particular, we first provide the research questions and the motivations behind their formulation. Then, data employed for the analysis is described, followed by the presentation of the validation methods.
In the last part of the section, the evaluation criteria we adopted for assessing the predictions achieved with the built machine learning models and threats to validity discussion are presented. 
The  data  and  scripts to train the models and reproduce the results may be found online at \url{https://tinyurl.com/US-privacy}.

\subsection{Research questions}\label{sec:research_objectives}
The aim of our investigation is to assess the application of advanced methods and technologies to detect privacy content from  USs.  As mentioned in the introduction, we have first performed a sort of sanity check to verify: 

\begin{itemize}
    \item [a)] if a deep learning method ($CNN_{NLP}$) performs at least as shallow machine learning  methods, when NLP-based features are exploited; 
    \item [b)] if a deep learning method ($CNN_{PW}$) performs at least as shallow machine learning  methods, when privacy word (PW) features are exploited.
\end{itemize}

Then, starting from the consideration that user story datasets are difficult to obtain,
especially those containing sensible information, we have investigated the use of Transfer Learning (TL) which allows developers to analyze the similarities between different tasks and to exploit a neural network used for one task in a given domain and apply it to another domain.

To conduct this research study, we have formulated  three research questions:

\begin{itemize}
    \item [\textbf{RQ1}] Is $CNN_{NLP}$ accurate at least as conventional machine learning methods to detect privacy content when using NLP-based features?
    \item [\textbf{RQ2}] Is $CNN_{PW}$ accurate at least as conventional machine learning methods to detect privacy content when using PW features?
     \item [\textbf{RQ3}] Are predictions obtained with $PD_{TL}$ better than those achieved with $CNN_{NLP}$ and $CNN_{PW}$?
\end{itemize}

To answer RQ1 we have considered a  convolutional network that is
trained on different features extracted through different NLP techniques 
to predict if USs contain privacy information. The considered neural networks are powerful and flexible models that have the ability to detect complex patterns even with limited training data. These models have showed high performance in several domains, including natural language processing \cite{Luong2015EffectiveAT}. It is therefore reasonable to assume that they are also effective in this context.
 
As for conventional machine learning methods, we have considered:   Logistic Regression (LR), Support Vector Machine (SVM), Gaussian Naive Bayes (GNB), k-Nearest Neighbors (kNN), Random Forest (RF), and Decision  Tree (DT). In the following, we name the models using NLP-based features as: $LR_{NLP}$, $SVM_{NLP}$, $GNB_{NLP}$, $kNN_{NLP}$, ${RF_{NLP}}$, and $DT_{NLP}$.
The choice of using approaches like LR and SVM is not accidental: they are often used in the literature for 
solving relevant problems in software engineering. Moreover, they are particularly suitable when dealing with binary classification tasks.

Similarly, for addressing RQ2 we built and compare models obtained with CNN, LR, SVM, GNB, kNN, RF, and DT using PWs as features.
In the following, we name these models as $CNN_{PW}$, $LR_{PW}$, $SVM_{PW}$, $GN_{BPW}$, $kNN_{PW}$, $RF_{PW}$, and $DT_{PW}$.

To address RQ3, we have considered 
the CNN, named $PD_{TL}$, defined in Section \ref{3.4}.
In particular, the expectation is that the model resulting from transfer learning can provide better predictions than $CNN_{NLP}$ and $CNN_{PW}$, as $CNN_{NLP}$ is trained on few data containing privacy information and does not exploit PW features, while $CNN_{PW}$ analyzes USs on a smaller set of features than $PD_{TL}$. 

\subsection{Data Collection}
\label{sec:data_collection}
The proposed model for the detection of privacy disclosures in USs needs data on which it is trained in order to make predictions. In particular, the data it needs should consist of a set of USs, each enriched by a label indicating whether that US has privacy disclosures, and by as many features as possible that contribute to the assertion of privacy relations. Datasets of this type, or similar, have not been found either on the Web or in the literature. Therefore, there was a need to build such a dataset, starting from a set of USs from which to extrapolate the characteristics that the model needs to make reliable predictions. To this end, a search was carried out to identify a large set of USs: this led to the discovery of 22 publicly available datasets, each containing more than 50 USs  \cite{Dalpiaz2018RequirementsDS}. 
The method used to obtain these datasets is described in detail in \cite{DalpiazSBAL19}.

Table \ref{table:overviewProjects} reports details and statistics about the considered datasets. Each row provides a brief description of the project, the number of USs, the number of privacy terms contained in the USs, and statistics about the NLP features used by the proposed approaches. In particular, each US was processed through the different NLP techniques in order to extrapolate the useful features to the subsequently defined models.
The last four columns of the table indicate the percentages of USs containing: both Privacy Words and Disclosures (\textit{PW\&Di}), only Privacy Words (\textit{PW}), only Disclosures (\textit{Di}), none of the above (\textit{None}). Note that the first author of the paper was in charge of  manually classifying the privacy information, while the other two cross-checked the data. 
Table  \ref{table:overviewDatasets} shows four USs together with the extracted NLP features.

\begin{table*}[width=\textwidth, t!]
\centering
 \caption{Properties of the datasets used for the research.}
\resizebox{\textwidth}{!}{%
  \begin{tabular}
  {>{\centering}m{0.8cm}p{9.3cm}>{\centering}m{0.5cm}>{\centering}m{0.9cm}p{1.2cm}p{1.2cm}p{1.2cm}p{1.2cm}}
    \hline
    \textbf{Dataset} & \textbf{Description} & \textbf{Size} & \multicolumn{1}{m{0.9cm}}{\textbf{Privacy
     Terms}}& \textbf{{\centering}\%PW\&Di}&\textbf{{\centering} \%PW}&\textbf{{\centering}\%Di}&\textbf{{\centering}\%None}\\
    \hline
    1 & Online platform for delivering transparent information on US governmental spending & 98 & 118 & \ChartPWDi{0.224} & \ChartPW{0.194} & \ChartDi{0.388} & \ChartNone{0.194}\\
    \hline
    2 & Electronic land management system for the Loudoun County, Virginia & 58 & 107 & \ChartPWDi{0.328} & \ChartPW{0.000} & \ChartDi{0.638} & \ChartNone{0.340}\\
    \hline
    3 & An online platform to support waste recycling & 51 & 86 & \ChartPWDi{0.176} & \ChartPW{0.137} & \ChartDi{0.137} & \ChartNone{0.549}\\
    \hline
    4 & Website for create a transparent overview of governmental expenses & 53 & 85 & \ChartPWDi{0.566} & \ChartPW{0.151} & \ChartDi{0.170} & \ChartNone{0.113}\\
    \hline
    5 & Platform for obtaining insights from data & 66 & 69 & \ChartPWDi{0.742} & \ChartPW{0.091} & \ChartDi{0.106} & \ChartNone{0.061}\\
    \hline
    6 & First version of the Scrum Alliance Website & 97 & 115 & \ChartPWDi{0.175} & \ChartPW{0.031} & \ChartDi{0.670} & \ChartNone{0.124}\\ 
    \hline
    7 & New version of the NSF website: redesign and content discovery & 73 & 115 & \ChartPWDi{0.041} & \ChartPW{0.000} & \ChartDi{0.740} & \ChartNone{0.219}\\ 
    \hline
    8 & App for camp administrators and parents & 55 & 56 & \ChartPWDi{0.273} & \ChartPW{0.182} & \ChartDi{0.164} & \ChartNone{0.382} \\ 
    \hline
    9 & First version of the PlanningPoker.com website & 53 & 53 & \ChartPWDi{0.170} & \ChartPW{0.057} & \ChartDi{0.623} & \ChartNone{0.151}\\ 
    \hline
    10 & Platform to find, share and publish data online & 67 & 63 & \ChartPWDi{0.552} & \ChartPW{0.134} & \ChartDi{0.104} & \ChartNone{0.209}\\ 
    \hline
    11 & Management information system for Duke University & 68 & 132 & \ChartPWDi{0.206} & \ChartPW{0.191} & \ChartDi{0.206} & \ChartNone{0.397}\\ 
    \hline
    12 & Simplified toolbox to enable fast and easy development with Hadoop & 64 & 67 & \ChartPWDi{0.109} & \ChartPW{0.219} & \ChartDi{0.219} & \ChartNone{0.453}\\ 
    \hline
    13 & Research data management portal for the university of Oxford, Reading and Southampton & 102 & 119 & \ChartPWDi{0.186} & \ChartPW{0.186} & \ChartDi{0.245} & \ChartNone{0.382}\\ 
    \hline
    14 & Personal interactive assistant for independent living and active aging & 138 & 126 & \ChartPWDi{0.036} & \ChartPW{0.065} & \ChartDi{0.413} & \ChartNone{0.486}\\ 
    \hline
    15 & Conference registration and management platform & 69 & 106 & \ChartPWDi{0.116} & \ChartPW{0.430} & \ChartDi{0.739} & \ChartNone{0.101}\\ 
    \hline
    16 & Software for machine-actionable data management plans & 83 & 115 & \ChartPWDi{0.578} & \ChartPW{0.181} & \ChartDi{0.229} & \ChartNone{0.012}\\ 
    \hline
    17 & Web-based archiving information system & 57 & 72 & \ChartPWDi{0.123} & \ChartPW{0.070} & \ChartDi{0.211} & \ChartNone{0.592}\\ 
    \hline
    18 & Institutional data repository for the University of Bath & 53 & 89 & \ChartPWDi{0.660} & \ChartPW{0.038} & \ChartDi{0.226} & \ChartNone{0.075}\\ 
    \hline
    19 & Repository for different types of digital content & 100 & 88 & \ChartPWDi{0.050} & \ChartPW{0.120} & \ChartDi{0.220} & \ChartNone{0.610}\\ 
    \hline
    20 & Software for archivists & 100 & 117 & \ChartPWDi{0.250} & \ChartPW{0.130} & \ChartDi{0.430} & \ChartNone{0.190} \\ 
    \hline
    21 & Digital content management system for Cornell University & 115 & 173 & \ChartPWDi{0.252} & \ChartPW{0.157} & \ChartDi{0.391} & \ChartNone{0.200}\\ 
    \hline
    22 & Citizen science platform that allows anyone to help in research tasks & 60 & 82 & \ChartPWDi{0.050} & \ChartPW{0.067} & \ChartDi{0.400} & \ChartNone{0.483}\\
    \hline
  \end{tabular}}
 \label{table:overviewProjects}
\end{table*}

\begin{table*}[width=1.9\linewidth,cols=8,pos=h]
\caption{Overview of the dataset used for the empirical study.}\label{table:overviewDatasets}
\begin{adjustbox}{angle=270}
\begin{tabular}{ p{0,4cm}|p{2,2cm}|p{3,5cm}|p{3,5cm}|p{4cm}|p{2,7cm}|p{1cm}|p{1,6cm} }
\toprule
\rowcolor{gray!10}  
\textbf{\#}& \textbf{User Story} & \textbf{Entities} & \textbf{Dependencies} & \textbf{Parts of Speech} & \textbf{Privacy Categories} & \textbf{Privacy Words} & \textbf{Disclosure?}\\
\hline
0 & 
    As a Data user, I want to have the 12-19-2017 deletions processed. & ['As', 'a', 'Data', 'user', 'PERSON', 'want', 'to', 'have', 'the', '12', '19', '2017', 'deletions', 'processed'] & 
    ['prep', 'det', 'compound', 'pobj', 'nsubj', 'ROOT', 'aux', 'xcomp', 'det', 'nummod', 'nummod', 'nummod', 'dobj', 'acl'] & 
    ['SCONJ', 'DET', 'PROPN', 'NOUN', 'PRON', 'VERB', 'PART', 'AUX', 'DET', 'NUM', 'NUM', 'NUM', 'NOUN', 'VERB'] & 
    [['PrivateSecret', 1]] & 
    ['data'] & 
    0 \\
    \hline
    1 & 
    As a UI designer, I want to redesign the Resources page, so that it matches the new Broker design styles. &
    ['As', 'a', 'HEALTH', 'HEALTH', 'PERSON', 'want', 'to', 'redesign', 'the', 'Resources', 'page', 'so', 'that', 'it', 'matches', 'the', 'new', 'PRODUCT', 'design', 'styles'] &
    ['prep', 'det', 'compound', 'pobj', 'nsubj', 'ROOT', 'aux', 'xcomp', 'det', 'compound', 'dobj', 'mark', 'mark', 'nsubj', 'advcl', 'det', 'amod', 'compound', 'compound', 'dobj'] &
    ['SCONJ', 'DET', 'PROPN', 'NOUN', 'PRON', 'VERB', 'PART', 'VERB', 'DET', 'PROPN', 'NOUN', 'SCONJ', 'SCONJ', 'PRON', 'VERB', 'DET', 'ADJ', 'PROPN', 'NOUN', 'NOUN'] &
    none &
    none &
    0 \\
    \hline
    2 &
    As a UI designer, I want to report to the Agencies about user testing, so that they are aware of their contributions to making Broker a better UX. &
    ['As', 'a', 'HEALTH', 'HEALTH', 'PERSON', 'want', 'to', 'report', 'to', 'the', 'ORG', 'about', 'user', 'testing', 'so', 'that', 'PERSON', 'are', 'aware', 'of', 'their', 'contributions', 'to', 'making', 'PRODUCT', 'a', 'better', 'UX'] &
    ['prep', 'det', 'compound', 'pobj', 'nsubj', 'ROOT', 'aux', 'xcomp', 'prep', 'det', 'pobj', 'prep', 'compound', 'pobj', 'mark', 'mark', 'nsubj', 'advcl', 'acomp', 'prep', 'poss', 'pobj', 'prep', 'pcomp', 'nsubj', 'det', 'amod', 'ccomp'] &
    ['SCONJ', 'DET', 'PROPN', 'NOUN', 'PRON', 'VERB', 'PART', 'VERB', 'ADP', 'DET', 'PROPN', 'ADP', 'NOUN', 'NOUN', 'SCONJ', 'SCONJ', 'PRON', 'AUX', 'ADJ', 'ADP', 'DET', 'NOUN', 'ADP', 'VERB', 'PROPN', 'DET', 'ADJ', 'PROPN'] &
    [['OpenVisible', 1]] &
    ['report'] &
    1\\
    \hline
    3 & 
    As a UI designer, I want to move on to round 2 of DABS or FABS landing page edits, so that I can get approvals from leadership. &
    ['As', 'a', 'HEALTH', 'HEALTH', 'PERSON', 'want', 'to', 'move', 'on', 'to', 'round', 'CARDINAL', 'of', 'DABS', 'or', 'FABS', 'landing', 'page', 'edits', 'so', 'that', 'PERSON', 'can', 'get', 'approvals', 'from', 'leadership'] &
    ['prep', 'det', 'compound', 'pobj', 'nsubj', 'ROOT', 'aux', 'xcomp', 'prt', 'aux', 'advcl', 'nummod', 'prep', 'pobj', 'cc', 'compound', 'compound', 'nsubj', 'conj', 'mark', 'mark', 'nsubj', 'aux', 'advcl', 'dobj', 'prep', 'pobj'] &
    ['SCONJ', 'DET', 'PROPN', 'NOUN', 'PRON', 'VERB', 'PART', 'VERB', 'ADV', 'PART', 'VERB', 'NUM', 'ADP', 'NOUN', 'CCONJ', 'NOUN', 'NOUN', 'NOUN', 'NOUN', 'SCONJ', 'SCONJ', 'PRON', 'VERB', 'AUX', 'NOUN', 'ADP', 'NOUN'] &
    none &
    none &
    1\\ 
\bottomrule
\end{tabular}
\end{adjustbox}
\end{table*}

\begin{figure}[t!]
\centering
\includegraphics[width=0.5\textwidth]{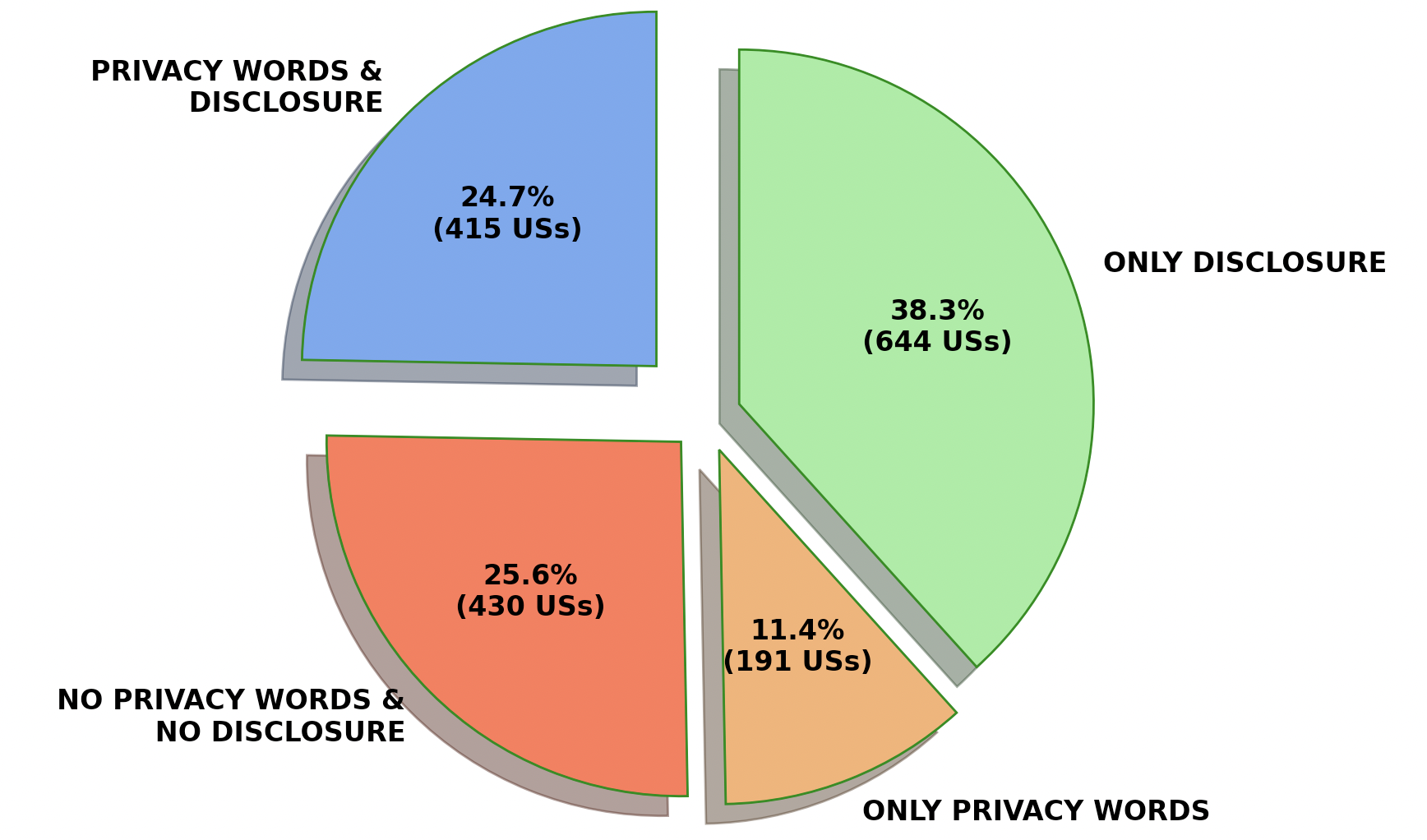}
\caption{Partitions of the dataset for each type of US.}
\label{fig:visual_data_1}
\end{figure}

This dataset was manually analyzed to verify if it was heterogeneous enough, i.e., if it included enough instances of each type of USs. The types of USs are the result of the assumption explained in the previous section. In particular, the types identified are: USs containing privacy words and disclosures, USs containing only privacy words, USs containing only disclosures, USs that do not contain neither privacy words nor disclosures. 
Figure \ref{fig:visual_data_1} shows the percentages of USs for each type for the considered dataset. Special attention was paid to types that contained only one of the two properties implying the presence of privacy content. If they did not have a large number of instances, there was a risk that the model would fail to differentiate correctly between the various types of USs, thus compromising the validity of the prediction.

The independent variables identified are the features extracted through NLP, thus entities, dependencies, parts of speech, privacy words, and privacy categories, while the dependent variables are Accuracy and F1-score. The choice of the latter variables is explained in the following section.

\subsection{Evaluation criteria}\label{sec:metrics}

To evaluate the accuracy of the predictions, we used four popular evaluation metrics for classification task \cite{metriche} – \textit{Accuracy}, \textit{Precision}, \textit{Recall}, and \textit{F1-score}. Accuracy is the most intuitive performance measure, and it is the ratio of the correctly predicted observations, i.e., \textit{true positive} + \textit{true negative}, to the total observations. 
Precision is calculated as \textit{true positive} / (\textit{true positive} +
\textit{false positive}) and indicates correctness of the responses provided by a technique. Recall measures the completeness of the responses and is calculated as \textit{true positive} / (\textit{true positive} + \textit{false negative}). F1-score is defined as the harmonic mean of precision and recall and indicates balance between those.

These types of evaluation metrics were firstly considered because in a binary classification task accuracy, precision and recall are equally important. Furthermore, these metrics allowed a comparison between the models implemented in this work and the pre-trained model evaluated on the same metrics.

The objective is to try to observe how precise the models are while identifying aspects of privacy, and to try to understand what are the limitations of their ability to extract as much as possible such aspects from the test dataset.

Furthermore, it was verified that the predictions obtained using the different models came from the same population in order to assess whether the differences observed by applying the chosen evaluation criterion (i.e., Accuracy and F1-score) were legitimate or due to coincidence \cite{doi:10.1080/00031305.2016.1154108}. 
Note that, non-parametric techniques are usually preferred \cite{articlegbmlri} to parametric methods when comparing machine learning and deep learning models mainly because they make fewer assumptions about the data. Thus, we decided to employ the McNemar test to compare the performance of two models \cite{d3979a14c7004e1f91b2862cf042799f, japkowicz_shah_2011}.
In particular, given the predictions of two models, A and B, and the truth labels, a contingency table is calculated, which examines the number of instances of the following: $i$) Both classifiers were correct; $ii$) 
Both classifiers were incorrect; $iii$) A was correct and B was incorrect; $iv$) B was correct and A was incorrect.
This makes it possible to estimate the probability that A is better than B at least as many times as observed in the experiment \cite{d3979a14c7004e1f91b2862cf042799f}. For comparing the performance of multiple machine learning and deep learning classifiers for the research questions in this thesis, the following null hypothesis was made:

\vspace{1mm}
\emph{Hn$_0$: All models are equally accurate in identifying aspects of privacy.}
\vspace{1mm}

McNemar's test allowed to test the null hypothesis by comparing each pair of models under the same null hypothesis. 
As usual we considered a \emph{p-value} of 0.05 as a ``significance'' threshold, i.e., \emph{p} values  lower than 0.05 are then assumed to be ``significant'', implying that the results obtained are hardly due to chance, allowing the null hypothesis to be rejected \cite{d3979a14c7004e1f91b2862cf042799f}. 
Thus, for the comparisons in which the null hypothesis was successfully rejected, it was determined whether one classifier was significantly better than the other classifiers.

\subsection{Validation Method}
\label{sec:valid_met}
In order to define the degree of accuracy or effectiveness of a machine learning model, one or more evaluations are carried out on the errors that are obtained in the predictions. In that case, after training, an error estimate is made for the model, called \textit{residual evaluation}. However, this estimate only gives an idea of how well the model does on the data used to train it, as it is possible for the model to be inadequate or in excess of the data. Thus, the problem with this evaluation technique is that it does not give an indication of how well the learning model will generalize to an independent or invisible dataset, i.e., on data it has not already seen. To this end, we have applied a k-fold cross-validation, dividing the original dataset into training and validation sets $k$ times, considering $k=5$. Furthermore, we run this 5-cross-validation 40 times. 
First, the cardinality of the sets defined by the assumption that determines whether the US is related to privacy content was analyzed. These cardinalities are shown in  Figure \ref{fig:visual_data_1}.

Using cross-validation, the sets defined for training consisted of 664 instances, where 50\% are USs containing both privacy words and disclosures, the remaining 50\% being divided between the other three types as described below. The test set was defined as 166 instances, where 83 consisted of USs containing both disclosures and privacy words.

\subsection{Threats to validity}\label{sec:threats}
This part discusses the main threats to validity, explaining their possible effect and how they have been mitigated. Threats to the validity of this work derive mainly from the correctness of the tools used, the assumption regarding privacy content, and the generalizability and repeatability of the  presented results.

\textbf{Construct Validity} It is about making sure that the measurement method corresponds to the construct being measured and is about the adequacy of the observations and inferences made based on the measurements performed during the study. In the context of using deep learning techniques for privacy content detection, methods offered by the Scikit-learn library have been used. In particular, the \verb|f1_score|\footnote[1]{\url{https://scikit-learn.org/stable/modules/generated/sklearn.metrics.f1\_score.html\#sklearn-metrics-f1-score}} method for measuring F1-score, and the \verb|accuracy_score|\footnote[2]{\url{https://scikit-learn.org/stable/modules/generated/sklearn.metrics.accuracy\_score.html}} method for measuring Accuracy were used from Scikit-learn. Relying on results from a single tool can pose a threat to validity especially in the case of deep-learning. However, here it was decided to choose Accuracy and F1-score as they were used in previous studies investigating an approach to detect privacy disclosures and this allowed the results obtained here to be compared with those obtained in those studies.
In addition, since the aim is to compare these deep learning techniques with classical machine learning models, the choice of these methods was almost obligatory, as the evaluation methods offered by Keras are not compatible with the Scikit-Learn models. On the other hand, the metrics offered by the latter library are based on results and predictions and, therefore, are also suitable for neural networks built using Keras.

\textbf{Internal Validity} It refers to the validity of the research results. It is mainly concerned with validating the control of extraneous variables and external influences that may impact the result. In the context of this work, exploring the applicability of transfer-learning for the detection of privacy aspects, it was assumed that the models used are compatible with each other, as they are both produced using the same technology, i.e., Keras. It would be interesting to observe how two models or neural networks developed through different APIs (e.g., Keras and Pytorch\footnote[1]{\url{https://pytorch.org/}}) can be combined in a transfer-learning experiment. Another threat to internal validity could be causality. However, it is assumed that the necessary conditions for causality are quite fulfilled as statistically significant correlations have been found between measures obtained via different methods, reinforcing the idea that these correlations derive from fairly robust causal relationships.

\textbf{External Validity} It relates to the generalizability and repeatability of the  produced results. The approach proposed in this work is based on Python. However, the statistical models used are replicable in other programming languages, so it is assumed that this method is programming language agnostic and therefore can be repeated for any other programming language given the availability of suitable frameworks. 
To promote the replication and construction of this work, as said above, we made available all tools, scripts and data.

\textbf{Conclusion Validity} It is a measure of how reasonable a research or experimental conclusion is. Although the number of observations made on statistical tests is not large, all the required hypotheses have been proved, therefore, the relationship between the data and the result is considered reasonable.

\section{Results and discussion}\label{sec:results}
In this section we present and discuss the results of the empirical study for each research question addressed.

\subsection{RQ1: Is $CNN_{NLP}$ accurate at least as conventional machine learning methods to detect privacy content when using NLP-based features?}
\label{sec:results_rq1}

 We present the results achieved with the models built to verify if a deep learning method (CNN) exploiting NLP-based features performs at least as conventional (shallow) machine  learning  methods (our first sanity check).  

As described in the previous Section \ref{sec:data_collection}, the built models are trained and tested with the same number of positive and negative samples. In particular, the positive samples are those with both Disclosures and Privacy Words, so for each fold 332 positive and 332 negative samples are selected for the training phase, while for the testing phase 83 positive and 83 negative samples are selected.

Each of the folds identified for the training was used for building $CNN_{NLP}$ (i.e., the model obtained with NLP-based CNN), $LR_{NLP}$, $SVM_{NLP}$, $GNB_{NLP}$, $kNN_{NLP}$, ${RF_{NLP}}$, and $DT_{NLP}$ (i.e., the models obtained with conventional machine learning methods by exploiting Scikit-Learn).

\begin{figure}[t!]
\centering
\includegraphics[width=0.475\textwidth]{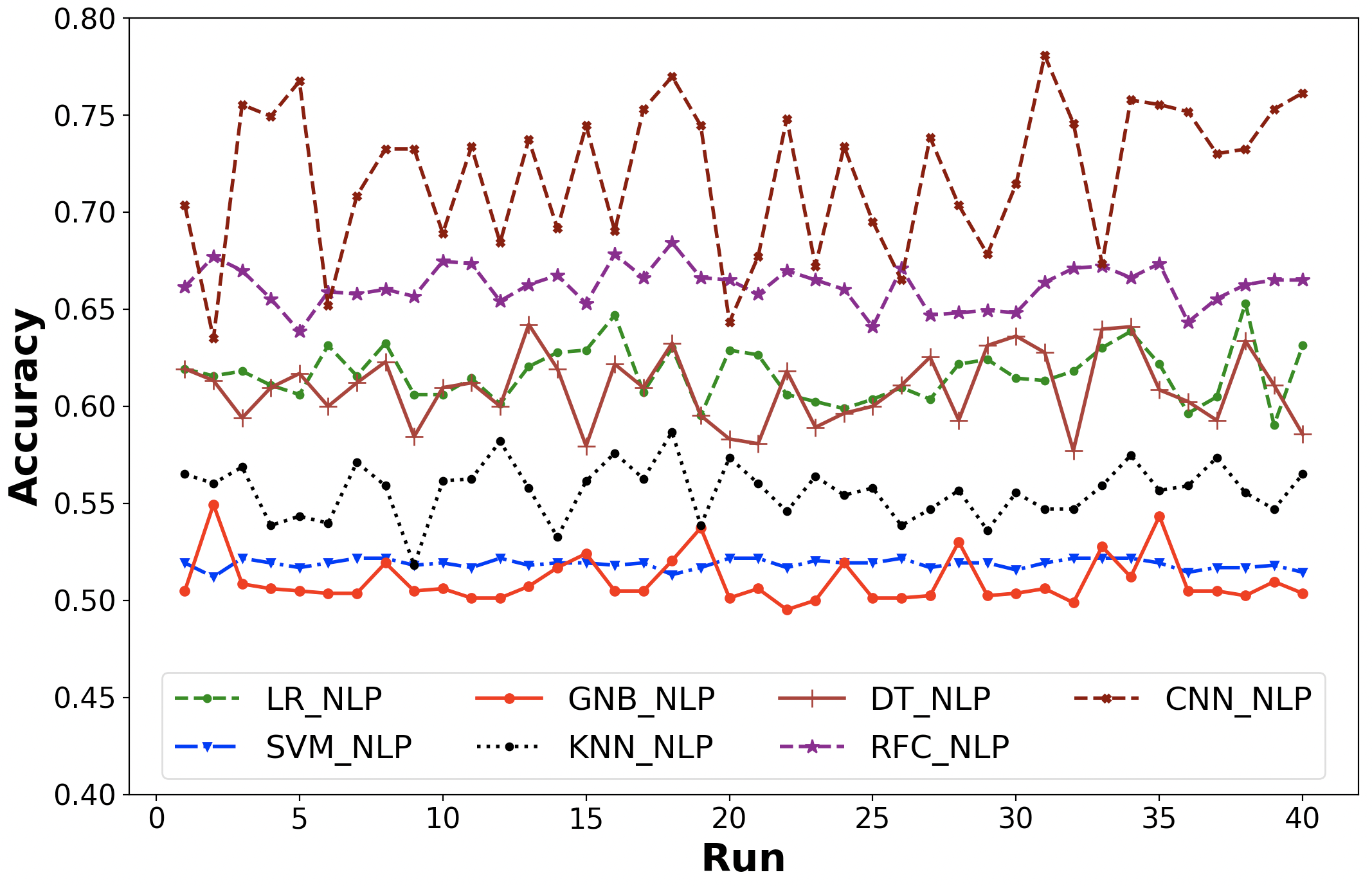}
\caption{Accuracy values of all the runs (to answer RQ1).}
\label{fig:accuracyRQ1}
\end{figure}

\begin{figure}[t!]
\centering
\includegraphics[width=0.475\textwidth]{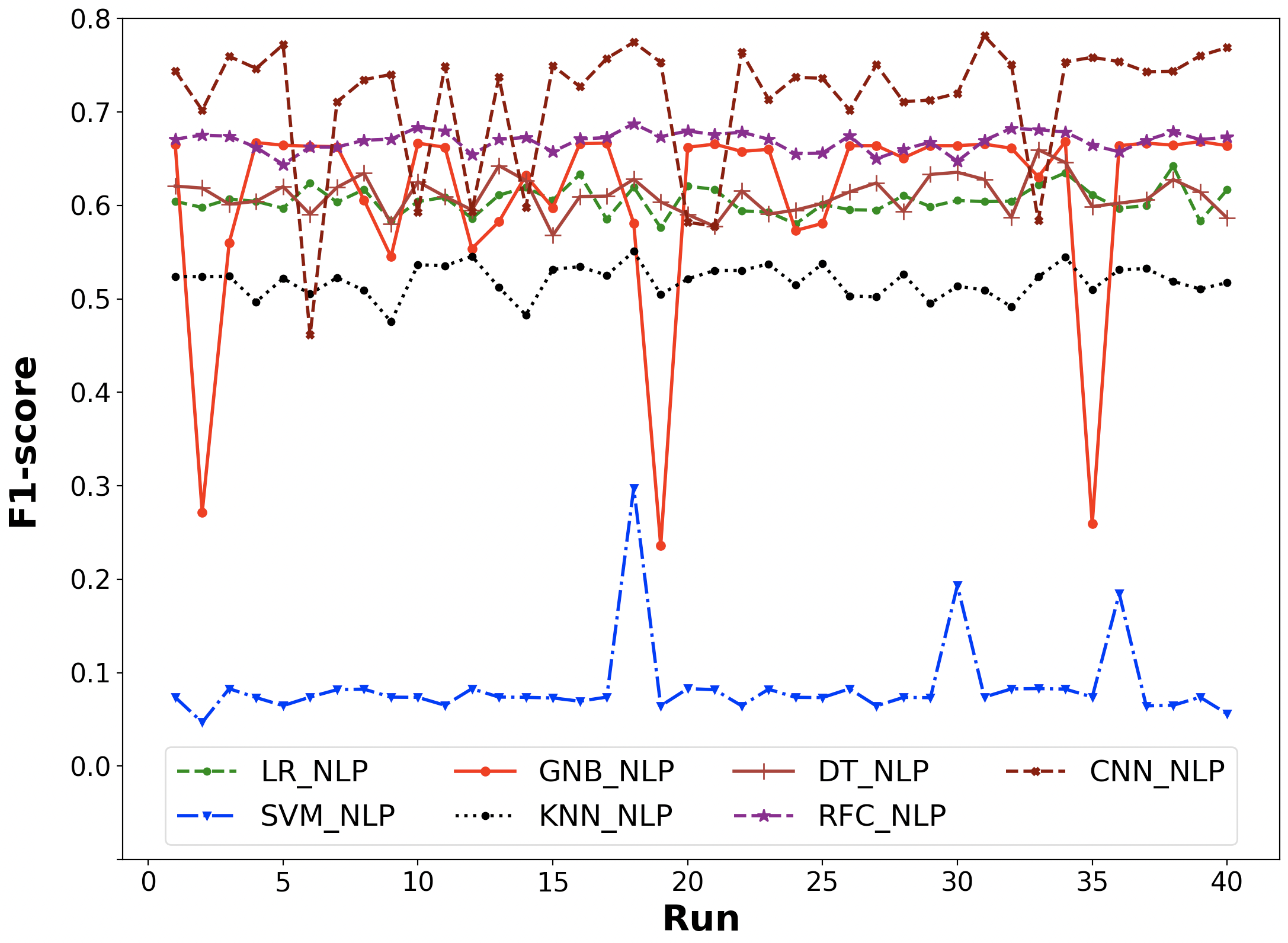}
\caption{F1-score values of all the runs (to answer RQ1).}
\label{fig:f1scoreRQ1}
\end{figure}

The aggregated results achieved in terms of the employed evaluation criteria are reported in Table \ref{table:results_logpwcnnRQ1}, while Figures \ref{fig:accuracyRQ1} and  \ref{fig:f1scoreRQ1} show the results of all runs graphically.
By analyzing the accuracy and F1-score values reported in Table \ref{table:results_logpwcnnRQ1}, we can observe that the results of 
$CNN_{NLP}$ are 
better  than those  obtained by the others. 
Indeed, accuracy and F1-score values are 0.720 and 0.713 for $CNN_{NLP}$, respectively, while the other machine learning methods have obtained values less than 0.7.
The worst results have been obtained with $SVM_{NLP}$. 

From Figure \ref{fig:accuracyRQ1} we can observe that $CNN_{NLP}$ is characterized by better Accuracy values for all runs, except for four cases. It is also interesting to note that for all methods we have a regular trend about the Accuracy values for all the runs, with a variation of up to 10\%. For a few cases we have variations around 20\% in the case of $CNN_{NLP}$. This is probably due to the syntactic structure of the USs selected for training phase. In particular, a greater number of positive and negative samples with a similar syntactic structure hampers the model to learn the presence of privacy aspects.

From Figure \ref{fig:f1scoreRQ1} we can observe that for $LR_{NLP}$, $KNN_{NLP}$, $DT_{NLP}$, and $RFC_{NLP}$ we have a regular trend about the F1-score values for all the runs (with a variation of up to 10\%). Differently, 
for $CNN_{NLP}$, $GNB_{NLP}$, and $SVM_{NLP}$ we can note some runs characterized by a variation around 20\%. Only in seven runs  $CNN_{NLP}$ is characterized by F1-score values below those of other approaches.  

\begin{table}[thb]
\caption{Results achieved with each model to answer RQ1 in terms of accuracy and F1-score.}\label{table:results_logpwcnnRQ1}
\begin{tabular*}{\tblwidth}{@{} CCC@{} }
\toprule
\rowcolor{gray!0}
\textbf{Model} & \textbf{Accuracy} & \textbf{F1-Score} \\
\midrule
$\mathbf{CNN_{NLP}}$  &  \textbf{0.720} & \textbf{0.713} \\
$LR_{NLP}$ & 0.617 & 0.605 \\
$SVM_{NLP}$ & 0.519 & 0.084\\
$GNB_{NLP}$ & 0.510 & 0.612\\
$kNN_{NLP}$ & 0.557 & 0.519\\
$RFC_{NLP}$ & 0.662 & 0.669\\
$DT_{NLP}$ & 0.609 & 0.611 \\
\bottomrule
\end{tabular*}
\end{table}

As designed we have also verified whether the  differences in the performances are statistically significant. To this end, we have performed the McNemar test to test the null hypothesis: ``there are no differences in the accuracy of the models being compared''. In particular, we have compared the predictions achieved with $CNN_{NLP}$ with those achieved with each shallow machine learning based model (i.e., those obtained with LR, SVM, GNB, kNN, RFC, and DT). 
For all the performed comparisons, we have obtained a \emph{p-value} \begin{math}<\end{math}0.001, allowing the rejection of the null hypothesis, i.e., there is significant differences between the predictions achieved with $CNN_{NLP}$ and those achieved with the employed shallow machine learning based model.
We can also conclude that the further effort needed to apply CNN is payed back by a significant improvement in the prediction accuracy.
\\

\begin{tabular}{|p{7cm}|}
\hline
Thus, we can positively answer our first research question because 
a deep learning method (CNN) has provided better predictions than conventional (shallow) machine learning methods.
\\
\hline
\end{tabular}

\subsection{RQ2 Is $CNN_{PW}$ accurate at least as conventional machine learning methods to detect privacy content when using PW features?}
\label{sec:results_rq2}
This section is devoted to the presentation of results achieved by models built to answer our second research question RQ2, i.e., if a deep learning method (CNN)  performs  at  least  as  shallow  machine  learning  methods, when PW features are exploited (our second sanity check).

Similarly to RQ1 analysis, the built models are trained and tested with the same number of positive and negative samples (see above). 
Thus, each of the folds identified for the training was used for the models built with CNN and LR, SVM, GNB, kNN, RF, and DT, by exploiting Scikit-Learn and PW features.

\begin{figure}[t!]
\centering
\includegraphics[width=0.475\textwidth]{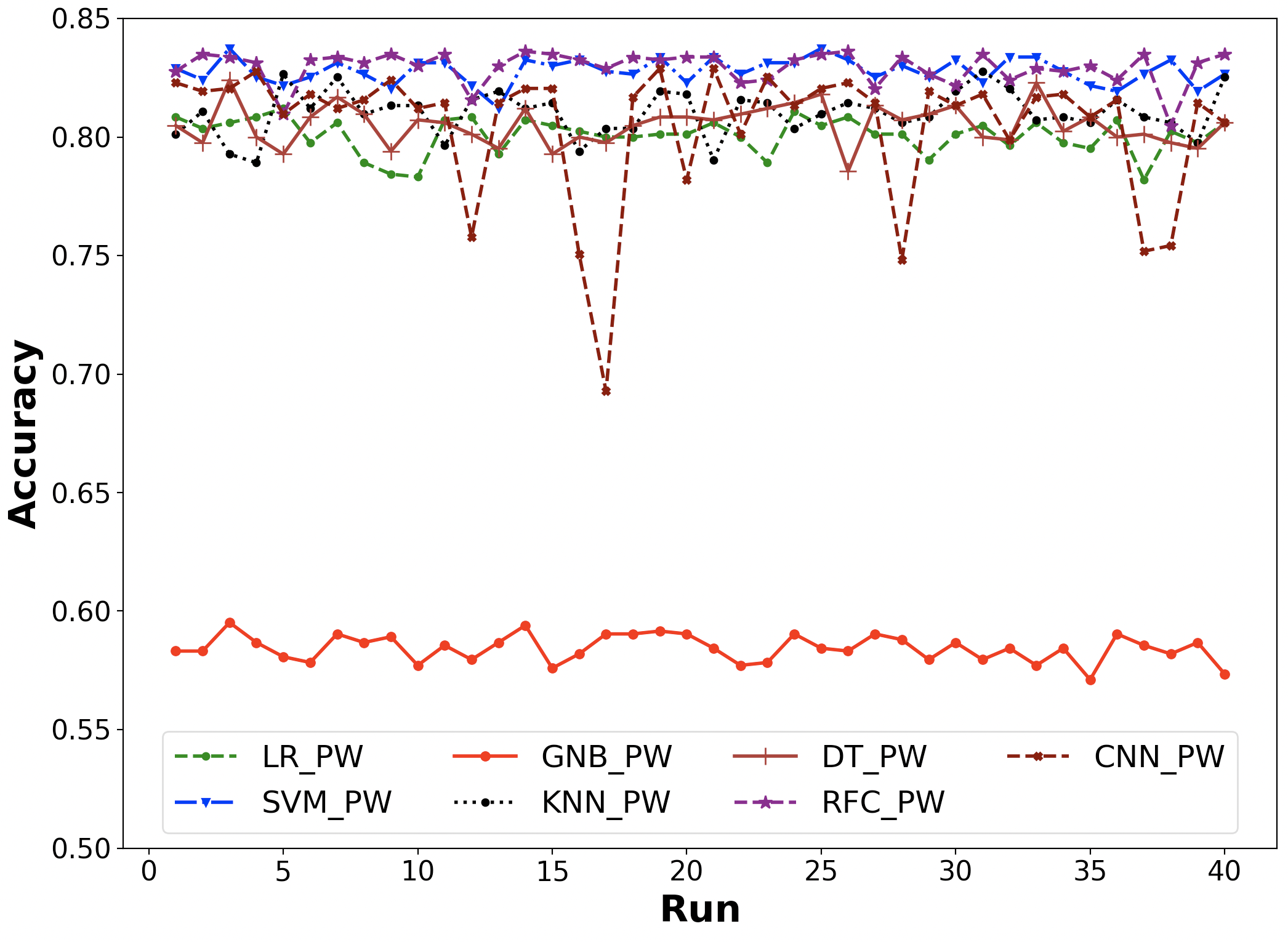}
\caption{Accuracy values of all the runs (to answer RQ2).}
\label{fig:accuracyRQ2}
\end{figure}

\begin{figure}[t!]
\centering
\includegraphics[width=0.4755\textwidth]{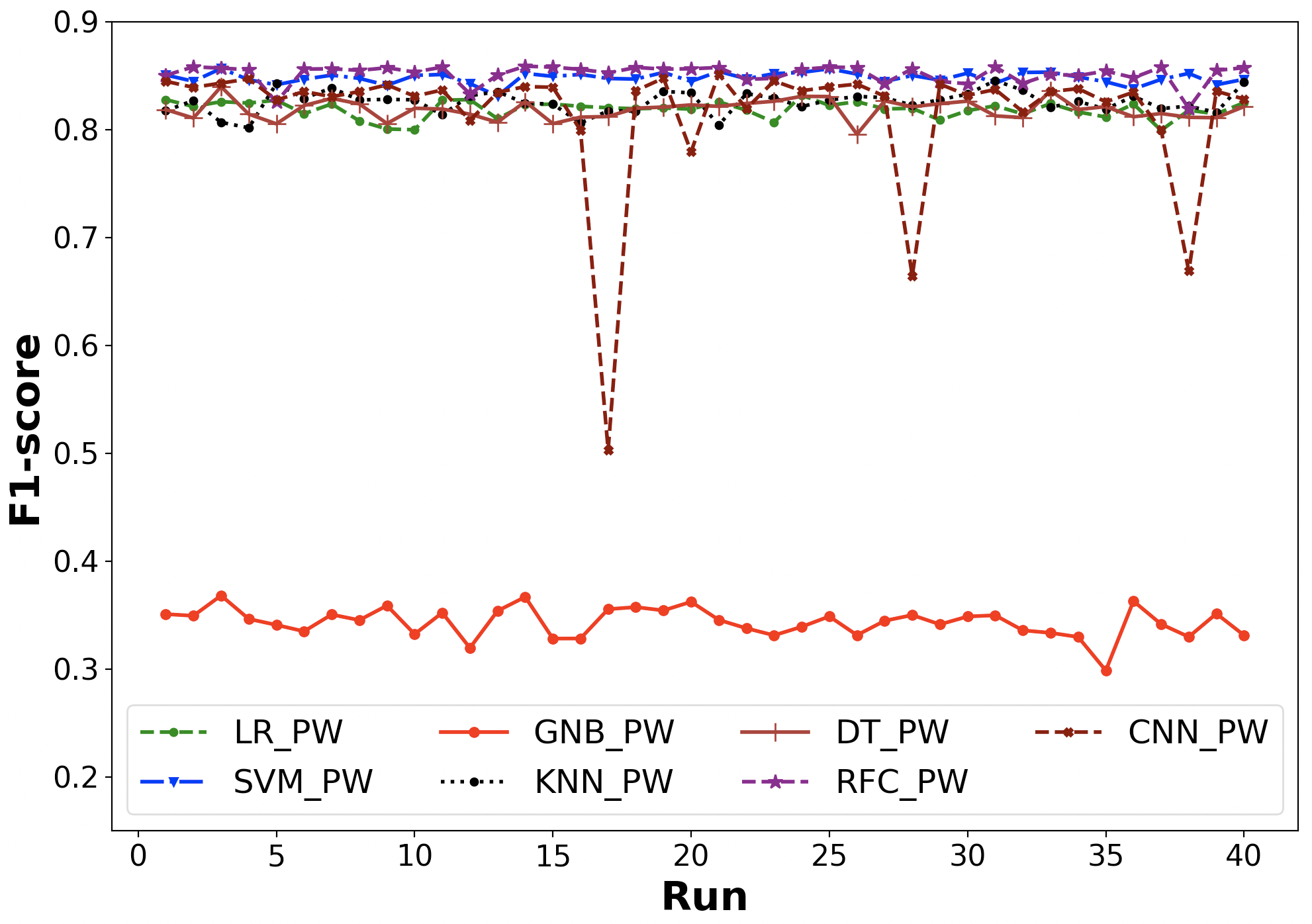}
\caption{F1-score values of all the runs (to answer RQ2).}
\label{fig:f1scoreRQ2}
\end{figure}

The aggregated results achieved in terms of employed evaluation criteria are reported in Table \ref{table:results_logpwcnn}, while Figures \ref{fig:accuracyRQ2} and  \ref{fig:f1scoreRQ2} show the results of all runs graphically.
By analyzing the Accuracy and F1-score values for the built models shown in Table \ref{table:results_logpwcnn}, we can observe that the values are  from 0.80 to 0.85, which can be considered good results, except for $GNB_{PW}$ which is characterized by worse performance with respect to the other employed machine learning methods. 

From Figures \ref{fig:accuracyRQ2} and  \ref{fig:f1scoreRQ2} we can observe that for all methods we have a regular trend for all the runs (with a variation of up to 10\%), except for $CNN_{PW}$ where for three runs the F1-score values are less than those of the other runs of about 20\%, and for just one run the Accuracy value is less than those of the other runs of about 14\%.
The best result in terms of accuracy and  F1-score has been obtained by using  $RF_{PW}$.  $SVM_{PW}$ and $kNN_{PW}$ have also provided better predictions than $CNN_{PW}$. They are reported in bold in Table \ref{table:results_logpwcnn}. 

As designed we have also verified whether the  differences in the performances are statistically significant, by performing the McNemar test.
For all the performed comparisons (i.e.,  $CNN_{PW}$ vs $LR_{PW}$, $CNN_{PW}$ vs $SVM_{PW}$, $CNN_{PW}$ vs $GNB_{PW}$, $CNN_{PW}$ vs $kNN_{PW}$, $CNN_{PW}$ vs $RF_{PW}$, $CNN_{PW}$ vs  $DT_{PW}$), we obtained a \emph{p-value} \begin{math}<\end{math}0.001, allowing the rejection of the null hypothesis, i.e., there is significant differences between the predictions achieved using the two considered models.
Thus,  $CNN_{PW}$ performs better than three conventional machine learning methods (i.e., $LR_{PW}$, $DT_{PW}$, and $GNB_{PW}$) and worse than the other three (i.e., $SVM_{PW}$, $kNN_{PW}$, and $RF_{PW}$), when  PW features are exploited.
\\   

\begin{tabular}{|p{7cm}|}
\hline
Thus, we cannot positively answer our second research question, i.e., the deep learning methods is not accurate at least as all the considered conventional machine learning methods to detect privacy content when using PW features.
\\ 
\hline
\end{tabular} \\

\color{black}

We can conclude that the second sanity check has been particularly useful because it  highlights something unexpected, i.e., a deep learning method is not accurate at least as a conventional machine learning method. But it can happen as shown in previous similar works (e.g., \cite{10.1093/jamia/ocz149}).   

Just for completeness, we want to observe that the prediction  models built with the shallow machine learning methods exploiting PW features are better than those obtained with the same  shallow machine learning methods but exploiting NLP-based features (see Tables  \ref{table:results_logpwcnnRQ1} and \ref{table:results_logpwcnn}). The results of McNemar test have also revealed that these differences are statistically significant.
Thus, the shallow machine learning methods improved their performances when trained with a not so large set of features. 

\begin{table}[thb]
\caption{Results achieved with each model to answer RQ2, in terms of accuracy and F1-score}
\label{table:results_logpwcnn}
\begin{tabular*}{\tblwidth}{@{} CCC@{} }
\toprule
\rowcolor{gray!0}
\textbf{Model} & \textbf{Accuracy} & \textbf{F1-Score} \\
\midrule
 $CNN_{PW}$  &  0.805 &  0.823 \\ 
 $LR_{PW}$ & 0.801  & 0.819\\
 $SVM_{PW}$ & \textbf{0.828}  & \textbf{0.848}\\
 $GNB_{PW}$ & 0.584  & 0.343\\
 $kNN_{PW}$ & \textbf{0.810}  & \textbf{0.825}\\ 
 $\mathbf{RF_{PW}}$ &  \textbf{0.829} & \textbf{0.851}\\
 $DT_{PW}$ & 0.805 & 0.819\\
\bottomrule
\end{tabular*}
\end{table}

\subsection{RQ3: Are predictions obtained with $PD_{TL}$ better than those achieved with $CNN_{NLP}$ and $CNN_{PW}$?}
\label{results_RQ3}
The main goal of our investigation is the attempt to apply the technique of Transfer Learning, which consists in using the knowledge of a model in solving a specific task and combining it with another model for solving a different task, expanding the set of features used for prediction.  

Similarly to RQ1 and RQ2 analyses, the comparisons between $PD_{TL}$, $CNN_{NLP}$, and $CNN_{PW}$ have been performed in terms of Accuracy and F1-score, whereas the McNemar's statistical test has been used to verify the significance of the achieved results.

\begin{figure}[t!]
\centering
\includegraphics[width=0.475\textwidth]{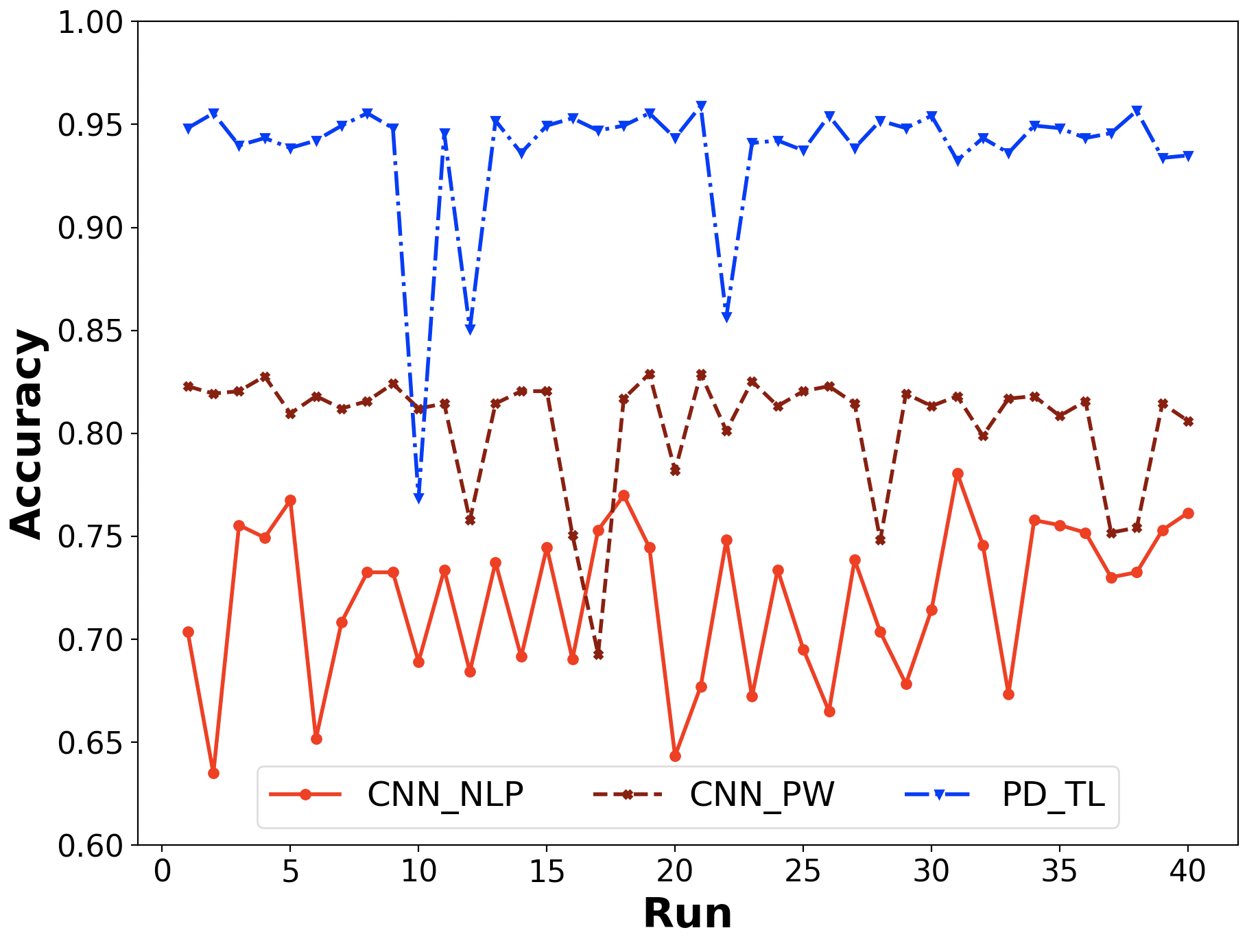}
\caption{Accuracy values of all the runs (to answer RQ3).}
\label{fig:accuracyRQ3}
\end{figure}

\begin{figure}[t!]
\centering
\includegraphics[width=0.475\textwidth]{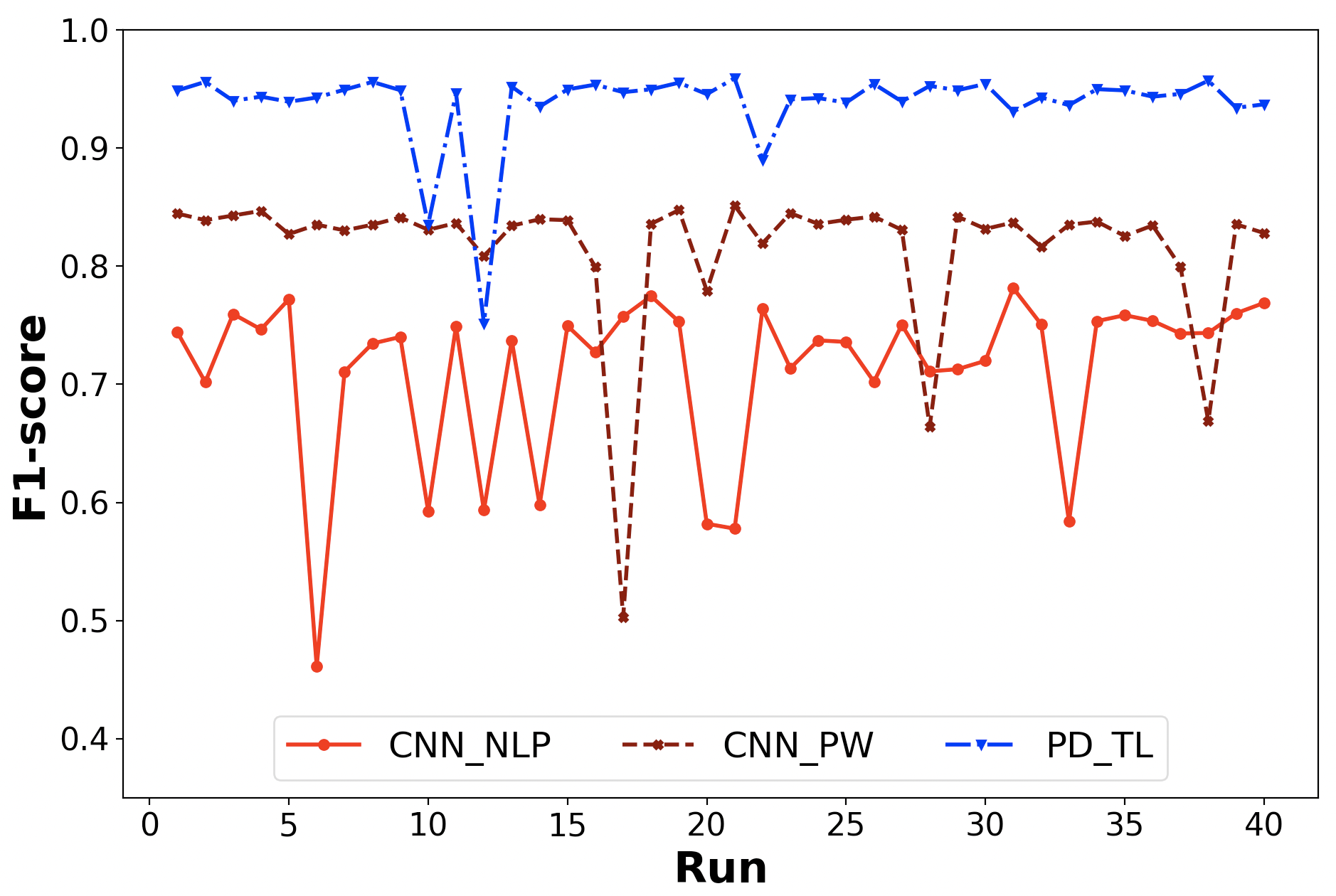}
\caption{F1-score values of all the runs (to answer RQ3).}
\label{fig:f1scoreRQ3}
\end{figure}

The aggregated results achieved in terms of employed evaluation criteria are reported in Table \ref{table:results_logpwcnnRQ3}, while Figures \ref{fig:accuracyRQ3}  and \ref{fig:f1scoreRQ3} show the results of all runs graphically. We can note that the model resulting from the application of the Transfer Learning ($PD_{TL}$) has provided better F1-score and Accuracy values (i.e., values greater than 0.90) than those achieved with the models based on deep learning  analyzed previously (i.e., $CNN_{NLP}$ and $CNN_{PW}$).
In particular,  $PD_{TL}$ surpasses $CNN_{PW}$ and $CNN_{NLP}$ of more than 10\% both in terms of Accuracy and F1-score.
Furthermore, the results of the McNemar test have revealed that the differences are statistically significant 
(p-value $<$0.001 for both the comparisons).
Furthermore, as clearly shown in Figures \ref{fig:accuracyRQ3} and  \ref{fig:f1scoreRQ3} $PD_{TL}$ has provided better results for all the runs except one, and the distribution of the values is characterized by less variation with respect to the ones of $CNN_{NLP}$ and $CNN_{PW}$.

\begin{table}[width=.9\linewidth,cols=4,pos=h]
\caption{Results achieved with each model to answer RQ3, in terms of accuracy and F1-score}
\label{table:results_logpwcnnRQ3}
\begin{tabular*}{\tblwidth}{@{} CCC@{} }
\toprule
\rowcolor{gray!0}
\textbf{Model} & \textbf{Accuracy} & \textbf{F1-Score} \\
\midrule
 $CNN_{NLP}$ & 0.720 & 0.713\\
 $CNN_{PW}$ & 0.805 & 0.823\\
 $\mathbf{PD_{TL}}$ & \textbf{0.937} & \textbf{0.937}\\ 
\bottomrule
\end{tabular*}
\end{table}

Based on the obtained results, it is therefore possible to state not only that Transfer Learning is feasible
but that it is better than using deep learning models alone for privacy content analysis.\\

\begin{tabular}{|p{7cm}|}
\hline
Thus, we can positively answer our third research question, i.e., predictions obtained with $PD_{TL}$ are better than those achieved with $CNN_{NLP}$ and $CNN_{PW}$.
\\
\hline
\end{tabular}

\subsection{Findings and  suggestions for researchers and practitioners}

The analysis carried out to answer our research questions allows us to highlight implications for researchers and practitioners about the applicability of our findings. We organize the discussion according to the achieved contributions. 

\smallskip
\begin{description}[leftmargin=0.3cm]

    \item[On the use of a tool to predict privacy content.] We have provided an approach and tool to automatically predict privacy content from user stories (problem never addressed before), which exploit a combination of NLP and transfer learning strategies. This should encourage software engineering researchers and in particular practitioners in considering the opportunities of automating privacy content detection.
    
 \noindent \faHandORight \xspace \underline{\emph{Implication 1.}} \emph{
 Practitioners have the possibility to exploit an approach and tool that allow to reduce the effort (and cost) to identify privacy requirements in the early phase of design. User studies involving practitioners should be performed  with the aim of promoting the suggested approach and tool.}

    \smallskip
  \item[On the use of deep learning methods.] As expected the  experimental results show that the use of NLP-based CNNs can contribute to improve predictions about privacy requirements with respect to the use of conventional (shallow) machine learning methods. However, the analysis has also revealed that the strategy for training the models are crucial. In particular, RQ2 analysis has not highlighted a clear advantage in using deep learning methods with respect to conventional (shallow) machine learning methods. Other studies achieved a similar result (e.g., \cite{10.1093/jamia/ocz149}).  
  
\smallskip
    \noindent \faHandORight \xspace \underline{\emph{Implication 2.}} \emph{Researchers should invest some effort in conducting empirical studies considering different datasets aiming at identifying strategies for training NLP-based prediction models in the context of agile for privacy requirement detection.}

  \smallskip
  \item[On the use of privacy words.] Our analysis has clearly shown that the use of privacy words allowed us to significantly improve the predictions of some employed shallow machine learning methods (if we compare RQ2 results against RQ1 results). In particular, $RF_{PW}$,  $SVM_{PW}$, and $kNN_{PW}$  have also provided better predictions than $CNN_{NLP}$. Thus, even cheaper methods can provide good predictions when exploiting  data of the specific domain under investigation.

\smallskip
    \noindent \faHandORight \xspace \underline{\emph{Implication 3.}} \emph{The research community should invest some effort in investigating the impact of the specific domain data on the use of cheaper methods aiming at verifying their effectiveness with respect to more expensive methods}.

\smallskip
    \item[On the use of Transfer Learning.] The main result of our analysis is about the use of Transfer Learning that has allowed us to improve the performance of the built NLP-based CNN prediction models of about 10\% in terms of ts). This is a further confirmation of the benefit of using this emergent strategy, which allows to reuse a system developed for a task to build a model for a different but related task \cite{torrey2010transfer,DBLP:journals/ese/KocaguneliMM15,DBLP:journals/tse/KrishnaM19}.  

  \noindent \faHandORight \xspace \underline{\emph{Implication 4.}} \emph{Researchers should apply Transfer Learning for training NLP-based prediction models aiming at improving their effectiveness in detecting privacy as well as security requirements in the agile context.}

\end{description}

\section{Conclusions and Future work} \label{sec:conclusions}
Interest in machine learning techniques based on natural language processing has been growing in recent years, including in the field of software engineering. Most of the existing attempts are focused on the generation of models and components useful in the different phases of software engineering from customer-specified requirements. On the other hand, few attempts to capture non-functional requirements have been documented in the literature, yet they contribute quite a bit in the evaluation of software quality. 

The results of our empirical study have revealed that deep learning methods can be used for the detection of non-functional requirements from customer requirements. In particular, it was found that deep learning models can be used for the identification of privacy disclosures in user stories, even with near-optimal performance. 
Furthermore, the search for recent deep learning techniques has led to the exploration of Transfer Learning, and therefore the possibility of its application in this context has been evaluated. The experiment on the application of Transfer Learning has demonstrated the feasibility of practicing this technique in the context of privacy content detection in user stories.


As for future research directions, there are reasons to extend this work to a broader scope, including other NFRs, or experimenting with such techniques for other similar tasks. Of course, future developments could also focus on improving the strategies employed in the work. For instance, an update of the privacy dictionary used or the production of newer, more elaborate taxonomies could help in this regard. Further research might involve the adoption of other NLP techniques for feature extraction, expanding the set on which the various models are trained and tested. Eventually, it would be interesting to analyze the application of Transfer Learning between models of different nature, both technological and methodological, in order to better understand in which contexts and circumstances this technique leads to significant improvements.

\printcredits



\bibliographystyle{IST-privacyUS}
\bibliography{IST-privacyUS}

\begin{thebibliography}{53}
\expandafter\ifx\csname natexlab\endcsname\relax\def\natexlab#1{#1}\fi
\providecommand{\url}[1]{\texttt{#1}}
\providecommand{\href}[2]{#2}
\providecommand{\path}[1]{#1}
\providecommand{\DOIprefix}{doi:}
\providecommand{\ArXivprefix}{arXiv:}
\providecommand{\URLprefix}{URL: }
\providecommand{\Pubmedprefix}{pmid:}
\providecommand{\doi}[1]{\href{http://dx.doi.org/#1}{\path{#1}}}
\providecommand{\Pubmed}[1]{\href{pmid:#1}{\path{#1}}}
\providecommand{\bibinfo}[2]{#2}
\ifx\xfnm\relax \def\xfnm[#1]{\unskip,\space#1}\fi
\bibitem[{Sommerville and Sawyer(1997)}]{Sommerville97RE}
\bibinfo{author}{I.~Sommerville}, \bibinfo{author}{P.~Sawyer},
  \bibinfo{title}{Requirements Engineering: A Good Practice Guide},
  \bibinfo{publisher}{Wiley}, \bibinfo{address}{New York, NY, USA},
  \bibinfo{year}{1997}.
\bibitem[{Pohl(2010)}]{10.5555/1869735}
\bibinfo{author}{K.~Pohl}, \bibinfo{title}{Requirements Engineering:
  Fundamentals, Principles, and Techniques}, \bibinfo{edition}{1st} ed.,
  \bibinfo{publisher}{Springer Publishing Company, Incorporated},
  \bibinfo{year}{2010}.
\bibitem[{Fernández et~al.(2017)Fernández, Wagner, Kalinowski, Felderer,
  Mafra, Vetrò, Conte, Christiansson, Greer, Lassenius, Männistö, Nayabi,
  Oivo, Penzenstadler, Pfahl, Prikladnicki, Ruhe, Schekelmann, Sen, Spinola,
  Tuzcu, de~la Vara, and Wieringa}]{twenty}
\bibinfo{author}{D.~M. Fernández}, \bibinfo{author}{S.~Wagner},
  \bibinfo{author}{M.~Kalinowski}, \bibinfo{author}{M.~Felderer},
  \bibinfo{author}{P.~Mafra}, \bibinfo{author}{A.~Vetrò},
  \bibinfo{author}{T.~Conte}, \bibinfo{author}{M.-T. Christiansson},
  \bibinfo{author}{D.~Greer}, \bibinfo{author}{C.~Lassenius},
  \bibinfo{author}{T.~Männistö}, \bibinfo{author}{M.~Nayabi},
  \bibinfo{author}{M.~Oivo}, \bibinfo{author}{B.~Penzenstadler},
  \bibinfo{author}{D.~Pfahl}, \bibinfo{author}{R.~Prikladnicki},
  \bibinfo{author}{G.~Ruhe}, \bibinfo{author}{A.~Schekelmann},
  \bibinfo{author}{S.~Sen}, \bibinfo{author}{R.~Spinola},
  \bibinfo{author}{A.~Tuzcu}, \bibinfo{author}{J.~L. de~la Vara},
  \bibinfo{author}{R.~Wieringa},
\newblock \bibinfo{title}{Naming the pain in requirements engineering -
  {C}ontemporary problems, causes, and effects in practice},
\newblock \bibinfo{journal}{Empirical software engineering}
  (\bibinfo{year}{2017}) \bibinfo{pages}{2298--2338}.
\bibitem[{Paetsch et~al.(2003)Paetsch, Eberlein, and
  Maurer}]{DBLP:conf/wetice/PaetschEM03}
\bibinfo{author}{F.~Paetsch}, \bibinfo{author}{A.~Eberlein},
  \bibinfo{author}{F.~Maurer},
\newblock \bibinfo{title}{Requirements engineering and agile software
  development},
\newblock in: \bibinfo{booktitle}{Proceedings of 12th {IEEE} International
  Workshops on Enabling Technologies {(WETICE} 2003), Infrastructure for
  Collaborative Enterprises, 9-11 June 2003, Linz, Austria},
  \bibinfo{publisher}{{IEEE} Computer Society}, \bibinfo{year}{2003}, pp.
  \bibinfo{pages}{308--313}. \DOIprefix\doi{10.1109/ENABL.2003.1231428}.
\bibitem[{{Kurtanović} and {Maalej}(2017)}]{RE2017}
\bibinfo{author}{Z.~{Kurtanović}}, \bibinfo{author}{W.~{Maalej}},
\newblock \bibinfo{title}{Automatically classifying functional and
  non-functional requirements using supervised machine learning},
\newblock in: \bibinfo{booktitle}{Proceedings of IEEE 25th International
  Requirements Engineering Conference (RE)}, \bibinfo{year}{2017}, pp.
  \bibinfo{pages}{490--495}. \DOIprefix\doi{10.1109/RE.2017.82}.
\bibitem[{Nguyen(2009)}]{DBLP:conf/icse/Nguyen09}
\bibinfo{author}{Q.~L. Nguyen},
\newblock \bibinfo{title}{Non-functional requirements analysis modeling for
  software product lines},
\newblock in: \bibinfo{booktitle}{Proceedings of {ICSE} Workshop on Modeling in
  Software Engineering, MiSE 2009, Vancouver, BC, Canada, May 17-18, 2009},
  \bibinfo{publisher}{{IEEE} Computer Society}, \bibinfo{year}{2009}, pp.
  \bibinfo{pages}{56--61}. \DOIprefix\doi{10.1109/MISE.2009.5069898}.
\bibitem[{{Slankas} and {Williams}(2013)}]{6611715}
\bibinfo{author}{J.~{Slankas}}, \bibinfo{author}{L.~{Williams}},
\newblock \bibinfo{title}{Automated extraction of non-functional requirements
  in available documentation},
\newblock in: \bibinfo{booktitle}{Proceedings of 1st International Workshop on
  Natural Language Analysis in Software Engineering (NaturaLiSE)},
  \bibinfo{year}{2013}, pp. \bibinfo{pages}{9--16}.
  \DOIprefix\doi{10.1109/NAturaLiSE.2013.6611715}.
\bibitem[{{Anthonysamy} et~al.(2017){Anthonysamy}, {Rashid}, and
  {Chitchyan}}]{7961663}
\bibinfo{author}{P.~{Anthonysamy}}, \bibinfo{author}{A.~{Rashid}},
  \bibinfo{author}{R.~{Chitchyan}},
\newblock \bibinfo{title}{Privacy requirements: Present future},
\newblock in: \bibinfo{booktitle}{Proceedings of IEEE/ACM 39th International
  Conference on Software Engineering: Software Engineering in Society Track
  (ICSE-SEIS)}, \bibinfo{year}{2017}, pp. \bibinfo{pages}{13--22}.
  \DOIprefix\doi{10.1109/ICSE-SEIS.2017.3}.
\bibitem[{Cao and Ramesh(2008)}]{nine}
\bibinfo{author}{L.~Cao}, \bibinfo{author}{B.~Ramesh},
\newblock \bibinfo{title}{Agile requirements engineering practices: An
  empirical study},
\newblock \bibinfo{journal}{IEEE Software}  (\bibinfo{year}{2008})
  \bibinfo{pages}{60--67}.
\bibitem[{Paetsch et~al.(2003)Paetsch, Eberlein, and Maurer}]{thirtyfour}
\bibinfo{author}{F.~Paetsch}, \bibinfo{author}{A.~Eberlein},
  \bibinfo{author}{F.~Maurer},
\newblock \bibinfo{title}{Requirements engineering and agile software
  development},
\newblock in: \bibinfo{booktitle}{Proceedings of IEEE International Workshops
  on Enabling Technologies: Infrastructure for Collaborative Enterprises},
  \bibinfo{year}{2003}, pp. \bibinfo{pages}{308--313}.
  \DOIprefix\doi{10.1109/ENABL.2003.1231428}.
\bibitem[{LeCun et~al.(2015)LeCun, Bengio, and Hinton}]{lecun2015deeplearning}
\bibinfo{author}{Y.~LeCun}, \bibinfo{author}{Y.~Bengio},
  \bibinfo{author}{G.~Hinton},
\newblock \bibinfo{title}{Deep learning},
\newblock \bibinfo{journal}{Nature} \bibinfo{volume}{521}
  (\bibinfo{year}{2015}) \bibinfo{pages}{436--444}.
\bibitem[{Mehdy et~al.(2019)Mehdy, Kennington, and
  Mehrpouyan}]{10.1007/978-3-030-21373-2_14}
\bibinfo{author}{N.~Mehdy}, \bibinfo{author}{C.~Kennington},
  \bibinfo{author}{H.~Mehrpouyan},
\newblock \bibinfo{title}{Privacy disclosures detection in natural-language
  text through linguistically-motivated artificial neural networks},
\newblock in: \bibinfo{booktitle}{Security and Privacy in New Computing
  Environments}, \bibinfo{year}{2019}, pp. \bibinfo{pages}{152--177}.
  \DOIprefix\doi{10.1007/978-3-030-21373-2\_14}.
\bibitem[{Li et~al.(2020)Li, Peng, Li, Xia, Yang, Sun, Yu, and
  He}]{DBLP:journals/corr/abs-2008-00364}
\bibinfo{author}{Q.~Li}, \bibinfo{author}{H.~Peng}, \bibinfo{author}{J.~Li},
  \bibinfo{author}{C.~Xia}, \bibinfo{author}{R.~Yang},
  \bibinfo{author}{L.~Sun}, \bibinfo{author}{P.~S. Yu},
  \bibinfo{author}{L.~He},
\newblock \bibinfo{title}{A survey on text classification: From shallow to deep
  learning},
\newblock \bibinfo{journal}{CoRR} \bibinfo{volume}{abs/2008.00364}
  (\bibinfo{year}{2020}).
\bibitem[{Haneczok and Piskorski(2020)}]{HANECZOK2020102371}
\bibinfo{author}{J.~Haneczok}, \bibinfo{author}{J.~Piskorski},
\newblock \bibinfo{title}{Shallow and deep learning for event relatedness
  classification},
\newblock \bibinfo{journal}{Information Processing \& Management}
  \bibinfo{volume}{57} (\bibinfo{year}{2020}) \bibinfo{pages}{102371}.
\bibitem[{Oleynik et~al.(2019)Oleynik, Kugic, Kasáč, and
  Kreuzthaler}]{10.1093/jamia/ocz149}
\bibinfo{author}{M.~Oleynik}, \bibinfo{author}{A.~Kugic},
  \bibinfo{author}{Z.~Kasáč}, \bibinfo{author}{M.~Kreuzthaler},
\newblock \bibinfo{title}{{Evaluating shallow and deep learning strategies for
  the 2018 n2c2 shared task on clinical text classification}},
\newblock \bibinfo{journal}{Journal of the American Medical Informatics
  Association} \bibinfo{volume}{26} (\bibinfo{year}{2019})
  \bibinfo{pages}{1247--1254}.
\bibitem[{{Xu} et~al.(2019){Xu}, {Qi}, {Yu}, {Xu}, {Zhao}, and
  {Yuan}}]{8945891}
\bibinfo{author}{G.~{Xu}}, \bibinfo{author}{C.~{Qi}},
  \bibinfo{author}{H.~{Yu}}, \bibinfo{author}{S.~{Xu}},
  \bibinfo{author}{C.~{Zhao}}, \bibinfo{author}{J.~{Yuan}},
\newblock \bibinfo{title}{Detecting sensitive information of unstructured text
  using convolutional neural network},
\newblock in: \bibinfo{booktitle}{Proceedings of International Conference on
  Cyber-Enabled Distributed Computing and Knowledge Discovery (CyberC)},
  \bibinfo{year}{2019}, pp. \bibinfo{pages}{474--479}.
  \DOIprefix\doi{10.1109/CyberC.2019.00087}.
\bibitem[{{Neerbeky} et~al.(2017){Neerbeky}, {Assentz}, and {Dolog}}]{7930091}
\bibinfo{author}{J.~{Neerbeky}}, \bibinfo{author}{I.~{Assentz}},
  \bibinfo{author}{P.~{Dolog}},
\newblock \bibinfo{title}{Taboo: Detecting unstructured sensitive information
  using recursive neural networks},
\newblock in: \bibinfo{booktitle}{Proceedings of IEEE 33rd International
  Conference on Data Engineering (ICDE)}, \bibinfo{year}{2017}, pp.
  \bibinfo{pages}{1399--1400}. \DOIprefix\doi{10.1109/ICDE.2017.195}.
\bibitem[{Torrey and Shavlik(2010)}]{torrey2010transfer}
\bibinfo{author}{L.~Torrey}, \bibinfo{author}{J.~Shavlik},
\newblock \bibinfo{title}{Transfer learning},
\newblock in: \bibinfo{booktitle}{Handbook of research on machine learning
  applications and trends: algorithms, methods, and techniques},
  \bibinfo{publisher}{IGI global}, \bibinfo{year}{2010}, pp.
  \bibinfo{pages}{242--264}.
\bibitem[{Kocaguneli et~al.(2015)Kocaguneli, Menzies, and
  Mendes}]{DBLP:journals/ese/KocaguneliMM15}
\bibinfo{author}{E.~Kocaguneli}, \bibinfo{author}{T.~Menzies},
  \bibinfo{author}{E.~Mendes},
\newblock \bibinfo{title}{Transfer learning in effort estimation},
\newblock \bibinfo{journal}{Empir. Softw. Eng.} \bibinfo{volume}{20}
  (\bibinfo{year}{2015}) \bibinfo{pages}{813--843}.
\bibitem[{Krishna and Menzies(2019)}]{DBLP:journals/tse/KrishnaM19}
\bibinfo{author}{R.~Krishna}, \bibinfo{author}{T.~Menzies},
\newblock \bibinfo{title}{Bellwethers: {A} baseline method for transfer
  learning},
\newblock \bibinfo{journal}{{IEEE} Trans. Software Eng.} \bibinfo{volume}{45}
  (\bibinfo{year}{2019}) \bibinfo{pages}{1081--1105}.
\bibitem[{Dalpiaz(2018)}]{Dalpiaz2018RequirementsDS}
\bibinfo{author}{F.~Dalpiaz}, \bibinfo{title}{Requirements data sets (user
  stories)}, \bibinfo{year}{2018}. \DOIprefix\doi{10.17632/7zbk8zsd8y.1},
  \bibinfo{note}{https://data.mendeley.com/datasets/7zbk8zsd8y/}.
\bibitem[{Baeza-Yates and Ribeiro-Neto(1999)}]{Baeza-Yates:1999}
\bibinfo{author}{R.~Baeza-Yates}, \bibinfo{author}{B.~Ribeiro-Neto},
  \bibinfo{title}{Modern Information Retrieval},
  \bibinfo{publisher}{Addison-Wesley}, \bibinfo{year}{1999}.
\bibitem[{Lucassen et~al.(2016)Lucassen, Dalpiaz, {van der Werf}, and
  Brinkkemper}]{formatus}
\bibinfo{author}{G.~Lucassen}, \bibinfo{author}{F.~Dalpiaz},
  \bibinfo{author}{J.~M. {van der Werf}}, \bibinfo{author}{S.~Brinkkemper},
\newblock \bibinfo{title}{The use and effectiveness of user stories in
  practice},
\newblock in: \bibinfo{editor}{M.~Daneva}, \bibinfo{editor}{O.~Pastor} (Eds.),
  \bibinfo{booktitle}{Requirements Engineering: Foundation for Software
  Quality}, \bibinfo{publisher}{Springer International Publishing},
  \bibinfo{address}{Cham}, \bibinfo{year}{2016}, pp. \bibinfo{pages}{205--222}.
\bibitem[{Cohn(2004)}]{cohnmodel}
\bibinfo{author}{M.~Cohn}, \bibinfo{title}{User Stories Applied: For Agile
  Software Development}, \bibinfo{publisher}{Addison Wesley},
  \bibinfo{year}{2004}.
\bibitem[{{Jiménez} and {Juárez-Ramírez}(2019)}]{9105527}
\bibinfo{author}{S.~{Jiménez}}, \bibinfo{author}{R.~{Juárez-Ramírez}},
\newblock \bibinfo{title}{A quality framework for evaluating grammatical
  structure of user stories to improve external quality},
\newblock in: \bibinfo{booktitle}{Proceedings of 7th International Conference
  in Software Engineering Research and Innovation (CONISOFT)},
  \bibinfo{year}{2019}, pp. \bibinfo{pages}{147--153}.
  \DOIprefix\doi{10.1109/CONISOFT.2019.00029}.
\bibitem[{{Lucassen} et~al.(2015){Lucassen}, {Dalpiaz}, {van der Werf}, and
  {Brinkkemper}}]{7320415}
\bibinfo{author}{G.~{Lucassen}}, \bibinfo{author}{F.~{Dalpiaz}},
  \bibinfo{author}{J.~M. {van der Werf}}, \bibinfo{author}{S.~{Brinkkemper}},
\newblock \bibinfo{title}{Forging high-quality user stories: Towards a
  discipline for agile requirements},
\newblock in: \bibinfo{booktitle}{Proceedings of IEEE 23rd International
  Requirements Engineering Conference (RE)}, \bibinfo{year}{2015}, pp.
  \bibinfo{pages}{126--135}. \DOIprefix\doi{10.1109/RE.2015.7320415}.
\bibitem[{Heck and Zaidman(2014)}]{heck2014quality}
\bibinfo{author}{P.~Heck}, \bibinfo{author}{A.~Zaidman}, \bibinfo{title}{A
  quality framework for agile requirements: A practitioner's perspective},
  \bibinfo{year}{2014}. \href{http://arxiv.org/abs/1406.4692}{\tt
  arXiv:1406.4692}.
\bibitem[{Karaa et~al.(2016)Karaa, Azzouz, Singh, Dey, Ashour, and
  Gh{\'{e}}zala}]{DBLP:journals/spe/KaraaASDAG16}
\bibinfo{author}{W.~B.~A. Karaa}, \bibinfo{author}{Z.~B. Azzouz},
  \bibinfo{author}{A.~Singh}, \bibinfo{author}{N.~Dey}, \bibinfo{author}{A.~S.
  Ashour}, \bibinfo{author}{H.~B. Gh{\'{e}}zala},
\newblock \bibinfo{title}{Automatic builder of class diagram {(ABCD):} an
  application of {UML} generation from functional requirements},
\newblock \bibinfo{journal}{Softw. Pract. Exp.} \bibinfo{volume}{46}
  (\bibinfo{year}{2016}) \bibinfo{pages}{1443--1458}.
\bibitem[{Elallaoui et~al.(2018)Elallaoui, Nafil, and Touahni}]{Procedia}
\bibinfo{author}{M.~Elallaoui}, \bibinfo{author}{K.~Nafil},
  \bibinfo{author}{R.~Touahni},
\newblock \bibinfo{title}{Automatic transformation of user stories into {UML}
  use case diagrams using {NLP} techniques},
\newblock \bibinfo{journal}{Procedia Computer Science} \bibinfo{volume}{130}
  (\bibinfo{year}{2018}) \bibinfo{pages}{42--49}. \bibinfo{note}{The 9th
  International Conference on Ambient Systems, Networks and Technologies (ANT
  2018) / The 8th International Conference on Sustainable Energy Information
  Technology (SEIT-2018) / Affiliated Workshops}.
\bibitem[{Nasiri et~al.(2020)Nasiri, Rhazali, Lahmer, and Chenfour}]{Procedia2}
\bibinfo{author}{S.~Nasiri}, \bibinfo{author}{Y.~Rhazali},
  \bibinfo{author}{M.~Lahmer}, \bibinfo{author}{N.~Chenfour},
\newblock \bibinfo{title}{Towards a generation of class diagram from user
  stories in agile methods},
\newblock \bibinfo{journal}{Procedia Computer Science} \bibinfo{volume}{170}
  (\bibinfo{year}{2020}) \bibinfo{pages}{831--837}. \bibinfo{note}{The 11th
  International Conference on Ambient Systems, Networks and Technologies (ANT)
  / The 3rd International Conference on Emerging Data and Industry 4.0 (EDI40)
  / Affiliated Workshops}.
\bibitem[{Lucassen et~al.(2017)Lucassen, Robeer, Dalpiaz, van~der Werf, and
  Brinkkemper}]{DBLP:journals/re/LucassenRDWB17}
\bibinfo{author}{G.~Lucassen}, \bibinfo{author}{M.~Robeer},
  \bibinfo{author}{F.~Dalpiaz}, \bibinfo{author}{J.~M. E.~M. van~der Werf},
  \bibinfo{author}{S.~Brinkkemper},
\newblock \bibinfo{title}{Extracting conceptual models from user stories with
  visual narrator},
\newblock \bibinfo{journal}{Requir. Eng.} \bibinfo{volume}{22}
  (\bibinfo{year}{2017}) \bibinfo{pages}{339--358}.
\bibitem[{{Robeer} et~al.(2016){Robeer}, {Lucassen}, {van der Werf}, {Dalpiaz},
  and {Brinkkemper}}]{7765525}
\bibinfo{author}{M.~{Robeer}}, \bibinfo{author}{G.~{Lucassen}},
  \bibinfo{author}{J.~M. {van der Werf}}, \bibinfo{author}{F.~{Dalpiaz}},
  \bibinfo{author}{S.~{Brinkkemper}},
\newblock \bibinfo{title}{Automated extraction of conceptual models from user
  stories via {NLP}},
\newblock in: \bibinfo{booktitle}{Proceedings of IEEE 24th International
  Requirements Engineering Conference (RE)}, \bibinfo{year}{2016}, pp.
  \bibinfo{pages}{196--205}. \DOIprefix\doi{10.1109/RE.2016.40}.
\bibitem[{{Gilson} and {Irwin}(2018)}]{8587287}
\bibinfo{author}{F.~{Gilson}}, \bibinfo{author}{C.~{Irwin}},
\newblock \bibinfo{title}{From user stories to use case scenarios towards a
  generative approach},
\newblock in: \bibinfo{booktitle}{Proceedings of 25th Australasian Software
  Engineering Conference (ASWEC)}, \bibinfo{year}{2018}, pp.
  \bibinfo{pages}{61--65}. \DOIprefix\doi{10.1109/ASWEC.2018.00016}.
\bibitem[{M{\"u}ter et~al.(2019)M{\"u}ter, Deoskar, Mathijssen, Brinkkemper,
  and Dalpiaz}]{Mter2019RefinementOU}
\bibinfo{author}{L.~M{\"u}ter}, \bibinfo{author}{T.~Deoskar},
  \bibinfo{author}{M.~Mathijssen}, \bibinfo{author}{S.~Brinkkemper},
  \bibinfo{author}{F.~Dalpiaz},
\newblock \bibinfo{title}{Refinement of user stories into backlog items:
  Linguistic structure and action verbs},
\newblock in: \bibinfo{editor}{E.~Knauss}, \bibinfo{editor}{M.~Goedicke}
  (Eds.), \bibinfo{booktitle}{Requirements Engineering: Foundation for Software
  Quality}, \bibinfo{publisher}{Springer International Publishing},
  \bibinfo{address}{Cham}, \bibinfo{year}{2019}, pp. \bibinfo{pages}{109--116}.
\bibitem[{Rane(2017)}]{Rane2017AutomaticGO}
\bibinfo{author}{P.~Rane}, \bibinfo{title}{Automatic Generation of Test Cases
  for Agile using Natural Language Processing}, Ph.D. thesis, Virginia Tech,
  \bibinfo{year}{2017}.
\bibitem[{{Gilson} et~al.(2019){Gilson}, {Galster}, and {Georis}}]{8712367}
\bibinfo{author}{F.~{Gilson}}, \bibinfo{author}{M.~{Galster}},
  \bibinfo{author}{F.~{Georis}},
\newblock \bibinfo{title}{Extracting quality attributes from user stories for
  early architecture decision making},
\newblock in: \bibinfo{booktitle}{Proceedings of IEEE International Conference
  on Software Architecture Companion (ICSA-C)}, \bibinfo{year}{2019}, pp.
  \bibinfo{pages}{129--136}. \DOIprefix\doi{10.1109/ICSA-C.2019.00031}.
\bibitem[{{Villamizar} et~al.(2019){Villamizar}, {Anderlin Neto}, {Kalinowski},
  {Garcia}, and {Méndez}}]{8920644}
\bibinfo{author}{H.~{Villamizar}}, \bibinfo{author}{A.~{Anderlin Neto}},
  \bibinfo{author}{M.~{Kalinowski}}, \bibinfo{author}{A.~{Garcia}},
  \bibinfo{author}{D.~{Méndez}},
\newblock \bibinfo{title}{An approach for reviewing security-related aspects in
  agile requirements specifications of web applications},
\newblock in: \bibinfo{booktitle}{Proceedings of IEEE 27th International
  Requirements Engineering Conference (RE)}, \bibinfo{year}{2019}, pp.
  \bibinfo{pages}{86--97}. \DOIprefix\doi{10.1109/RE.2019.00020}.
\bibitem[{Riaz et~al.(2014)Riaz, King, Slankas, and Williams}]{6912260}
\bibinfo{author}{M.~Riaz}, \bibinfo{author}{J.~King},
  \bibinfo{author}{J.~Slankas}, \bibinfo{author}{L.~Williams},
\newblock \bibinfo{title}{Hidden in plain sight: Automatically identifying
  security requirements from natural language artifacts},
\newblock in: \bibinfo{booktitle}{Proceedings of IEEE 22nd International
  Requirements Engineering Conference (RE)}, \bibinfo{year}{2014}, pp.
  \bibinfo{pages}{183--192}. \DOIprefix\doi{10.1109/RE.2014.6912260}.
\bibitem[{Barker et~al.(2009)Barker, Askari, Banerjee, Ghazinour, Mackas,
  Majedi, Pun, and Williams}]{10.1007/978-3-642-02843-4_7}
\bibinfo{author}{K.~Barker}, \bibinfo{author}{M.~Askari},
  \bibinfo{author}{M.~Banerjee}, \bibinfo{author}{K.~Ghazinour},
  \bibinfo{author}{B.~Mackas}, \bibinfo{author}{M.~Majedi},
  \bibinfo{author}{S.~Pun}, \bibinfo{author}{A.~Williams},
\newblock \bibinfo{title}{A data privacy taxonomy},
\newblock in: \bibinfo{booktitle}{Proceedings of the 26th British National
  Conference on Databases: Dataspace: The Final Frontier}, BNCOD 26,
  \bibinfo{publisher}{Springer-Verlag}, \bibinfo{address}{Berlin, Heidelberg},
  \bibinfo{year}{2009}, p. \bibinfo{pages}{42–54}.
  \DOIprefix\doi{10.1007/978-3-642-02843-4\_7}.
\bibitem[{De~Capitani Di~Vimercati et~al.(2012)De~Capitani Di~Vimercati,
  Foresti, Livraga, and Samarati}]{Vimercati12dataprivacy}
\bibinfo{author}{S.~De~Capitani Di~Vimercati}, \bibinfo{author}{S.~Foresti},
  \bibinfo{author}{G.~Livraga}, \bibinfo{author}{P.~Samarati},
\newblock \bibinfo{title}{Data privacy: Definitions and techniques},
\newblock \bibinfo{journal}{International Journal of Uncertainty, Fuzziness and
  Knowledge-Based Systems} \bibinfo{volume}{20} (\bibinfo{year}{2012})
  \bibinfo{pages}{793--817}.
\bibitem[{Gill et~al.(2011)Gill, Vasalou, Papoutsi, and
  Joinson}]{10.1145/1978942.1979421}
\bibinfo{author}{A.~J. Gill}, \bibinfo{author}{A.~Vasalou},
  \bibinfo{author}{C.~Papoutsi}, \bibinfo{author}{A.~N. Joinson},
\newblock \bibinfo{title}{Privacy dictionary: {A} linguistic taxonomy of
  privacy for content analysis},
\newblock in: \bibinfo{booktitle}{Proceedings of the SIGCHI Conference on Human
  Factors in Computing Systems}, \bibinfo{publisher}{ACM},
  \bibinfo{address}{New York, NY, USA}, \bibinfo{year}{2011}, p.
  \bibinfo{pages}{3227–3236}. \DOIprefix\doi{10.1145/1978942.1979421}.
\bibitem[{Vasalou et~al.(2011)Vasalou, Gill, Mazanderani, Papoutsi, and
  Joinson}]{PrivacyDictionary}
\bibinfo{author}{A.~Vasalou}, \bibinfo{author}{A.~Gill},
  \bibinfo{author}{F.~Mazanderani}, \bibinfo{author}{C.~Papoutsi},
  \bibinfo{author}{A.~Joinson},
\newblock \bibinfo{title}{Privacy dictionary: {A} new resource for the
  automated content analysis of privacy},
\newblock \bibinfo{journal}{Journal of the American Society for Information
  Science and Technology (JASIST)} \bibinfo{volume}{62} (\bibinfo{year}{2011})
  \bibinfo{pages}{2095--2105}.
\bibitem[{{Silva} et~al.(2020){Silva}, {Gonçalves}, {Godinho}, {Antunes}, and
  {Curado}}]{9162683}
\bibinfo{author}{P.~{Silva}}, \bibinfo{author}{C.~{Gonçalves}},
  \bibinfo{author}{C.~{Godinho}}, \bibinfo{author}{N.~{Antunes}},
  \bibinfo{author}{M.~{Curado}},
\newblock \bibinfo{title}{Using {NLP} and machine learning to detect data
  privacy violations},
\newblock in: \bibinfo{booktitle}{Proceedings of IEEE Conference on Computer
  Communications Workshops}, \bibinfo{year}{2020}, pp.
  \bibinfo{pages}{972--977}.
  \DOIprefix\doi{10.1109/INFOCOMWKSHPS50562.2020.9162683}.
\bibitem[{{Tesfay} et~al.(2019){Tesfay}, {Serna}, and {Rannenberg}}]{8931855}
\bibinfo{author}{W.~B. {Tesfay}}, \bibinfo{author}{J.~{Serna}},
  \bibinfo{author}{K.~{Rannenberg}},
\newblock \bibinfo{title}{Privacybot: Detecting privacy sensitive information
  in unstructured texts},
\newblock in: \bibinfo{booktitle}{Proceedings of Sixth International Conference
  on Social Networks Analysis, Management and Security (SNAMS)},
  \bibinfo{year}{2019}, pp. \bibinfo{pages}{53--60}.
  \DOIprefix\doi{10.1109/SNAMS.2019.8931855}.
\bibitem[{Sheth et~al.(2014)Sheth, Kaiser, and
  Maalej}]{10.1145/2568225.2568244}
\bibinfo{author}{S.~Sheth}, \bibinfo{author}{G.~Kaiser},
  \bibinfo{author}{W.~Maalej},
\newblock \bibinfo{title}{Us and them: A study of privacy requirements across
  north america, asia, and europe},
\newblock in: \bibinfo{booktitle}{Proceedings of the 36th International
  Conference on Software Engineering}, ICSE 2014,
  \bibinfo{publisher}{Association for Computing Machinery},
  \bibinfo{address}{New York, NY, USA}, \bibinfo{year}{2014}, p.
  \bibinfo{pages}{859–870}. \DOIprefix\doi{10.1145/2568225.2568244}.
\bibitem[{Evans and Zhai(1996)}]{evans-zhai-1996-noun}
\bibinfo{author}{D.~A. Evans}, \bibinfo{author}{C.~Zhai},
\newblock \bibinfo{title}{Noun phrase analysis in unrestricted text for
  information retrieval},
\newblock in: \bibinfo{booktitle}{Proceedings of 34th Annual Meeting of the
  Association for Computational Linguistics}, \bibinfo{publisher}{ACL},
  \bibinfo{address}{Santa Cruz, California, USA}, \bibinfo{year}{1996}, pp.
  \bibinfo{pages}{17--24}. \URLprefix \url{https://aclanthology.org/P96-1003}.
  \DOIprefix\doi{10.3115/981863.981866}.
\bibitem[{Luong et~al.(2015)Luong, Pham, and Manning}]{Luong2015EffectiveAT}
\bibinfo{author}{T.~Luong}, \bibinfo{author}{H.~Pham}, \bibinfo{author}{C.~D.
  Manning},
\newblock \bibinfo{title}{Effective approaches to attention-based neural
  machine translation},
\newblock \bibinfo{journal}{ArXiv} \bibinfo{volume}{abs/1508.04025}
  (\bibinfo{year}{2015}).
\bibitem[{Dalpiaz et~al.(2019)Dalpiaz, {Van Der Schalk}, Brinkkemper, Aydemir,
  and Lucassen}]{DalpiazSBAL19}
\bibinfo{author}{F.~Dalpiaz}, \bibinfo{author}{I.~{Van Der Schalk}},
  \bibinfo{author}{S.~Brinkkemper}, \bibinfo{author}{F.~B. Aydemir},
  \bibinfo{author}{G.~Lucassen},
\newblock \bibinfo{title}{Detecting terminological ambiguity in user stories:
  Tool and experimentation},
\newblock \bibinfo{journal}{Inf. Softw. Technol.} \bibinfo{volume}{110}
  (\bibinfo{year}{2019}) \bibinfo{pages}{3--16}.
\bibitem[{Powers(2011)}]{metriche}
\bibinfo{author}{D.~M.~W. Powers},
\newblock \bibinfo{title}{Evaluation: from precision, recall and f-measure to
  {ROC}, informedness, markedness and correlation},
\newblock \bibinfo{journal}{Journal of Machine Learning Technologies}
  \bibinfo{volume}{2} (\bibinfo{year}{2011}) \bibinfo{pages}{37--63}.
\bibitem[{Wasserstein and Lazar(2016)}]{doi:10.1080/00031305.2016.1154108}
\bibinfo{author}{R.~L. Wasserstein}, \bibinfo{author}{N.~A. Lazar},
\newblock \bibinfo{title}{The {ASA} statement on p-values: Context, process,
  and purpose},
\newblock \bibinfo{journal}{The American Statistician} \bibinfo{volume}{70}
  (\bibinfo{year}{2016}) \bibinfo{pages}{129--133}.
\bibitem[{Fernández et~al.(2011)Fernández, García, Luengo, Bernadó, and
  Herrera}]{articlegbmlri}
\bibinfo{author}{A.~Fernández}, \bibinfo{author}{S.~García},
  \bibinfo{author}{J.~Luengo}, \bibinfo{author}{E.~Bernadó},
  \bibinfo{author}{F.~Herrera},
\newblock \bibinfo{title}{Genetics-based machine learning for rule induction:
  State of the art, taxonomy, and comparative study},
\newblock \bibinfo{journal}{IEEE Transactions on Evolutionary Computation}
  \bibinfo{volume}{14} (\bibinfo{year}{2011}) \bibinfo{pages}{913 -- 941}.
\bibitem[{Salzberg(1997)}]{d3979a14c7004e1f91b2862cf042799f}
\bibinfo{author}{S.~Salzberg},
\newblock \bibinfo{title}{On comparing classifiers: Pitfalls to avoid and a
  recommended approach},
\newblock \bibinfo{journal}{Data Mining and Knowledge Discovery}
  \bibinfo{volume}{1} (\bibinfo{year}{1997}) \bibinfo{pages}{317--328}.
\bibitem[{Japkowicz and Shah(2011)}]{japkowicz_shah_2011}
\bibinfo{author}{N.~Japkowicz}, \bibinfo{author}{M.~Shah},
  \bibinfo{title}{Evaluating Learning Algorithms: A Classification
  Perspective}, \bibinfo{publisher}{Cambridge University Press},
  \bibinfo{year}{2011}. \DOIprefix\doi{10.1017/CBO9780511921803}.

\end{thebibliography}





\end{document}